\documentclass[pra,aps,superbib,citeautoscript,twocolumn]{revtex4-1}
\usepackage[english]{babel}
\usepackage[utf8]{inputenc}
\usepackage[T1]{fontenc}

\usepackage[colorlinks=true,urlcolor=blue,citecolor=blue,linkcolor=blue]{hyperref}

\usepackage{amsmath,bm}
\usepackage{amstext}
\usepackage{epsfig}
\usepackage{xcolor}
\usepackage{subfig}
\usepackage{graphicx,amsmath,amssymb,tabularx}
\usepackage{multirow}
\usepackage{array}
\usepackage{dsfont}
\usepackage{caption}
\usepackage{cleveref}
\usepackage[toc]{appendix}
\usepackage{xr}
\usepackage{physics}

\definecolor{dgreen}{rgb}{0,.5,0}
\definecolor{dred}{rgb}{.7,.0,.0}
\newcommand{\Eq}[1]{Eq.~(\ref{#1})}

\newcommand{\etal}{{\it et al.}}

\def\ddroit{{d}}
\def\im{{\rm i}}
\DeclareMathOperator*{\argmin}{arg\,min}
\DeclareMathAlphabet\mathbfcal{OMS}{cmsy}{b}{n}
\newcommand{\manu}[1]{{\textcolor{dgreen}{[ Manu: #1 ]}} }

\newcommand{\be}{\begin{eqnarray}}
\newcommand{\ee}{\end{eqnarray}}
\DeclareMathAlphabet\mathbfcal{OMS}{cmsy}{b}{n}

\begin{document}

\title{
Site-occupation--Green's function embedding theory: 
A density-functional approach to dynamical impurity solvers% for strongly correlated electrons
%towards an accurate treatment of strongly correlated systems without impurity solvers
%Merging site-occupation and Green's function embedding theories: towards an accurate treatment of strongly correlated systems without impurity solvers
}

\author{
Laurent Mazouin$^1$, Matthieu Saubanère$^2$, and Emmanuel Fromager$^1$
%, Vincent Robert and Masahisa Tsuchiizu 
}

\affiliation{\it 
~\\
$^1$Laboratoire de Chimie Quantique,\\
Institut de Chimie, CNRS / Universit\'{e} de Strasbourg,\\
1 rue Blaise Pascal, F-67000 Strasbourg, France\\
$^2$Institut Charles Gerhardt, CNRS/Universit\'{e} de Montpellier,\\ 
Place Eug\`{e}ne Bataillon, F-34095 Montpellier, France
}

%%%%% Abstract %%%%%

\begin{abstract}

%Site-occupation Green's function embedding theory (SOGET) is the reformulation of 
A reformulation of site-occupation embedding theory
(SOET) in terms of Green's functions is presented. Referred to as
site-occupation--Green's function embedding theory (SOGET), this novel extension of
density-functional theory for model Hamiltonians 
shares many
features with dynamical mean-field theory (DMFT) but is formally
exact (in any dimension). In SOGET, the impurity-interacting correlation
potential becomes a density-functional self-energy which is
frequency-dependent and in principle non-local. 
A simple local density-functional approximation (LDA) combining the Bethe Ansatz
(BA) LDA with the self-energy of the two-level Anderson model is
constructed and successfully applied to the one-dimensional
Hubbard model. Unlike in previous implementations of SOET, no many-body
wavefunction is needed, thus reducing drastically the computational
cost of the method. 
\end{abstract}

\maketitle

%\tableofcontents{}

%%%%% Article %%%%%

\section{Introduction}\label{sec:intro}

%\manu{TO DO LIST\\
%(1) check Eqs. numbering in Fig. 1\\
%(2) include updated refs from .bbl file\\
%(3) check new fig 2
%(4) new fig. 3 uploaded ? 
%}

The description of strong
electron correlation is a long-standing problem in both quantum
chemistry and condensed matter physics. In the former case,  
state-of-the-art {\it ab initio} methods are 
based on the explicit calculation of a many-body wavefunction.
Unfortunately, they can only be applied to relatively small systems
because of the exponentially increasing size of the many-body Hilbert
space. An in-principle-exact alternative is density-functional theory
(DFT)~\cite{hohenberg_inhomogeneous_1964,kohn_self-consistent_1965}
which drastically reduces the cost by mapping the fully interacting
system onto a non-interacting one. The bottleneck of DFT is, in
practice, the lack of accurate density-functional approximations that
can properly treat strongly correlated
systems~\cite{cohen_challenges_2012,swart_spin_2013,swart_spinning_2016}.
In the face of these problems, new methods have been developed. Due to
the fact that strong electron correlation is mainly local, only a
reduced part of the system has to be treated accurately. Hence, quantum embedding
methods~\cite{sun_quantum_2016,muhlbach_quantum_2018} have been gaining an increasing attention. They
deliver a good compromise between the accuracy of wavefunction-based
methods and the computational cost of mean-field-like methods like
DFT.\\ 

Turning to model Hamiltonians like Hubbard, the Green's function-based
dynamical mean-field theory
(DMFT)~\cite{georges_hubbard_1992,georges_dynamical_1996,kotliar_strongly_2004,held_electronic_2007,zgid_dynamical_2011}
is exact in the infinite dimension limit where the self-energy is momentum-independent and local. When merged with {\it ab initio} approaches like DFT~\cite{kotliar_electronic_2006} or GW~\cite{sun_extended_2002,biermann_first-principles_2003,karlsson_self-consistent_2005,boehnke_when_2016,werner_dynamical_2016,nilsson_multitier_2017},
the method can be applied to realistic systems. DMFT has been succesfull
in describing materials with localized $d$  and $f$ bands
\cite{haule_correlated_2008,miyake_d-_2008}. An appealing extension of Green's function techniques to quantum
chemical problems is self-energy embedding theory (SEET)~
\cite{kananenka_systematically_2015,lan_communication:_2015,zgid_finite_2017,lan_generalized_2017}
where local and non-local electronic correlations are modelled by the combination of wave function-based methods with many-body perturbation
theory. A mixture of configuration interaction with Green's
functions has also been recently proposed by Dvorak and
Rinke~\cite{dvorak_quantum_2019}.\\

For the calculation of non-dynamical properties, density matrix embedding
theory (DMET)~
\cite{knizia_density_2012,knizia_density_2013,bulik_density_2014,zheng_ground-state_2016,wouters_practical_2016,rubin_hybrid_2016,welborn_bootstrap_2016,wouters_five_2017,ye_incremental_2018,fertitta_rigorous_2018},
or the related {\it rotationally invariant slave bosons}
technique~\cite{ayral_dynamical_2017,lee_rotationally_2019}, has become
a viable alternative to DMFT. In standard DMET, the embedding procedure
relies on the Schmidt decomposition of a mean-field many-body wavefunction. The one-electron reduced density matrix is then introduced in
order to define a convergence criterion for the method. Note that DMET has also
been extended to the calculation of spectral
properties~\cite{booth_spectral_2015}.\\
 
Turning now to DFT for model Hamiltonians, also referred to as
site-occupation functional theory (SOFT)~\cite{capelle_density_2013}, a
local density approximation (LDA) based on the Bethe ansatz (BA) has
been developed for the one-dimensional (1D)
Hubbard model~\cite{lima_density-functional_2002,lima_density_2003,capelle_density-functional_2003}.
It contains the effect of strong correlation and can describe Mott
physics~\cite{silva_effects_2005,akande_electric_2010}. An
extension of BALDA to higher dimensions has been recently
proposed~\cite{vilela_approximate_2019}. 
Note also that the use of other
(frequency-independent) reduced
quantities, such as the one-body density matrix, has been considered in
the so-called lattice
DFT~\cite{lopez-sandoval_density-matrix_2002,lopez-sandoval_interaction-energy_2004,saubanere_scaling_2009,tows_lattice_2011,saubanere_density-matrix_2011,tows_density-matrix_2013,saubanere_interaction-energy_2016}.
In recent years, an
in-principle-exact alternative formulation of SOFT, referred to as
\textit{site-occupation embedding theory}
(SOET)~\cite{fromager_exact_2015,senjean_local_2017,senjean_site-occupation_2018,senjean_multiple_2018},
has been explored. In contrast to standard Kohn--Sham (KS) DFT, SOET maps
the whole physical system onto an impurity-interacting one. Note that
both systems have the {\it same} size. So far, the impurity-interacting system
has been treated by exact diagonalization for small
rings~\cite{senjean_local_2017} or on the level of density matrix
renormalization group
(DMRG)~\cite{nakatani_https://github.com/naokin_nodate,senjean_site-occupation_2018}
for (slightly)
larger systems. More recently, Senjean has formulated a projected
version of SOET, where the Schmidt decomposition is applied to the KS
determinant, thus reducing
drastically the computational cost of the method~\cite{senjean_projected_2019}. In order to access physical properties such as double
occupations and per-site energies, a density-functional correction is applied to the ``bare'' properties of the auxiliary impurity-interacting
system~\cite{senjean_site-occupation_2018}. The formal advantages of SOET over other hybrid methods is the absence
of double counting, the existence of a variational principle, as
well as in-principle-exact expressions for the physical properties of
interest. 
%Moreover, in the absence of proper density-functional approximations, the number of impurity sites can be increased in order to better reproduce physical quantities solely from the impurity-interacting system.
From a practical point of view, the major drawback of SOET is the size of
the impurity-interacting system which is not effectively reduced as in
DMET and makes calculations prohibitively expensive.\\ 

As mentioned
previously, a reduction in system size can be obtained by projection~\cite{senjean_projected_2019}, in
the spirit of DMET. In this work, we explore an alternative approach
%solution of this problem is projected SOET, where the degrees of freedom of the impurity-interacting system are drastically reduced via the Schmidt decomposition and only a small fragment is treated either by exact diagonalization or DMRG. 
based on the reformulation of SOET in terms of Green's functions. 
The approach will be referred to as \textit{site-occupation--Green's
function embedding theory} (SOGET) in the following. 
Instead of combining static
density-functional approximations with a many-body wavefunction
treatment, we introduce in SOGET a density-functional self-energy which is both
frequency- {\it and} site-occupation-dependent. The role of this
self-energy is to generate an impurity Green's function that
reproduces, in principle exactly, the site occupations of the physical
system. It also describes electron correlation in the auxiliary 
impurity-interacting system. In this work, we develop a simple LDA based on the combination of BALDA with the Anderson dimer model.
As our self-energy depends explicitly on the impurity site occupation, there
is no need for an impurity solver, thus reducing drastically the
computational cost of the method.\\ 

The paper is organized as follows. After a brief review on SOET
(Sec.~\ref{subsec:soet}), a formally exact derivation of SOGET is
presented in Sec.~\ref{subsec:soget}. The importance of derivative
discontinuities in the bath, 
when it comes to model gap openings,
is then
highlighted in Sec.~\ref{subsec:gap_DD_spec_fun}. In order to turn SOGET into a
practical computational method, density-functional approximations to
both correlation energies and the impurity-interacting self-energy must be
developed, as discussed in detail in Sec.~\ref{sec:approximations}. A
summary of the various approximations as well as computational details are
given in Sec.~\ref{sec:comp_details}. Results obtained with SOGET for
the 1D Hubbard model are presented and discussed in Sec.~\ref{sec:res_disc}.
Conclusions and perspectives are finally given in Sec.~\ref{sec:conc_persp}.

\section{Theory}\label{sec:theo}

\subsection{Site-occupation embedding theory}\label{subsec:soet}

The single-impurity version of SOET, which is considered in the rest of
this work, is briefly reviewed in the following. More details
can be found in Refs.~\cite{senjean_site-occupation_2018,senjean_multiple_2018}. 
%\subsubsection{Exact theory}\label{subsubsec:exact_soet}
% Hubbard model part
We start from the (grand canonical) 1D Hubbard Hamiltonian,
\begin{eqnarray}
\hat{H} = \hat{T} + \hat{U} + \hat{V} - \mu\hat{N},
\label{eq:hubbard_model}
\end{eqnarray}
where
\begin{eqnarray}
\hat{T} = -t\displaystyle{\sum_{<i,j>,\sigma}} (\hat{a}_{i\sigma}^{\dagger}\hat{a}_{j\sigma} + \mathrm{h.c.})
\label{eq:kinetic_operator}
\end{eqnarray}
describes the nearest-neighbor hopping of the electrons. It is
analogous to the kinetic energy operator in DFT. The parameter $t$ is
the hopping integral. The summation over site indices goes from $i=0$ to
$L-1$. In order to uniquely define the ground state, we impose
anti-periodic boundary conditions ($\hat{a}_{L\sigma} =
-\hat{a}_{0\sigma}$) when the number of electrons is a multiple of four and periodic boundary conditions ($\hat{a}_{L\sigma} = \hat{a}_{0\sigma}$) otherwise. The Coulomb operator
\begin{eqnarray}
\hat{U} = U\displaystyle{\sum_{i}}\hat{n}_{i\uparrow}\hat{n}_{i\downarrow}
\label{eq:coulomb_operator}
\end{eqnarray}
describes the on-site repulsion of electrons with interaction strength $U$. In addition to the kinetic and Coulomb term, we add a local potential operator
\begin{eqnarray}
\hat{V} = \sum_{i}v_{i}\hat{n}_{i}
\label{eq:local_potential_operator}
\end{eqnarray}
which plays the role of the external potential in DFT and therefore
modulates the electronic density (the occupation of the sites in this
context). The density operator on site $i$,
\begin{eqnarray}
\hat{n}_{i} = \sum_{\sigma}\hat{a}_{i\sigma}^{\dagger}\hat{a}_{i\sigma},
\label{eq:counting_operator}
\end{eqnarray}
yields the individual site occupations which can be viewed as a proxy of the electronic density in atomic systems. The chemical potential $\mu$ fixes the total number of electrons in the system via the particle number operator
\begin{eqnarray}
\hat{N} = \sum_{i}\hat{n}_{i} \, .
\label{eq:particle_number_operator}
\end{eqnarray}

% SOFT part
In the language of DFT~\cite{hohenberg_inhomogeneous_1964}, the exact ground-state energy is obtained
variationally, and for a
given number of electrons, as follows:
\begin{eqnarray}
E(\mathbf{v}) = \min_{\mathbf{n}}\lbrace F(\mathbf{n}) +
(\mathbf{v}|\mathbf{n}) \rbrace,
\label{eq:gs_energy}
\end{eqnarray}
where $\mathbf{n}\equiv(n_0,n_1,\ldots,n_{L-1})$ is the density profile, 
\begin{eqnarray}
F(\mathbf{n}) = \min_{\Psi\rightarrow\mathbf{n}}\langle\Psi | \big( \hat{T} + \hat{U} \big) | \Psi\rangle
\label{eq:levy_lieb_functional}
\end{eqnarray}
is the Levy-Lieb (LL) functional, and
\begin{eqnarray}
(\mathbf{v}|\mathbf{n}) = \sum_{i}v_{i}n_{i} \,.
\label{eq:scalar_product}
\end{eqnarray}
Within the conventional KS formalism, the expression in
Eq.~(\ref{eq:levy_lieb_functional}) is split as follows: 
\begin{eqnarray}
F(\mathbf{n}) = T_{\rm s}(\mathbf{n}) + E_{\rm Hxc}(\mathbf{n}), 
\label{eq:ks_partition}
\end{eqnarray}
where 
\begin{eqnarray}
T_{\rm s}(\mathbf{n}) = \min_{\Psi\rightarrow\mathbf{n}}\langle\Psi | \hat{T} |\Psi\rangle
\label{eq:ts_functional}
\end{eqnarray}
is the kinetic energy functional of the fictitious non-interacting
KS system with density $\mathbf{n}$ and 
\begin{eqnarray}
E_{\rm Hxc}(\mathbf{n}) = \dfrac{U}{4}\sum_{i}n_{i}^{2} + E_{\rm c}(\mathbf{n})
\label{eq:hxc_functional}
\end{eqnarray}
is the Hartree-exchange-correlation (Hxc) functional. In the particular
case of a uniform system, which is considered in the rest of this work,
the following local density approximation to the correlation energy
becomes exact:
\be
E_{\rm c}(\mathbf{n})=\sum_ie_{\rm c}(n_i)
,
\ee
where $e_{\rm c}(n)$ is the per-site density-functional correlation
energy.\\ 

% SOET part
In SOET, a different partitioning of the LL functional is used,   
\begin{eqnarray}
F(\mathbf{n}) = F^{\rm imp}(\mathbf{n}) + \overline{E}_{\rm Hxc}^{\rm
bath}(\mathbf{n}), 
\label{eq:soet_partition}
\end{eqnarray}
where 
\begin{eqnarray}
F^{\rm imp}(\mathbf{n}) &=&  \min_{\Psi\rightarrow\mathbf{n}}\langle\Psi |
\big( \hat{T} + \hat{U}_0) |\Psi\rangle,
\nonumber\\
&=&\left\langle\Psi^{\rm imp}(\mathbf{n})\middle\vert\hat{T} + \hat{U}_0\middle\vert \Psi^{\rm
imp}(\mathbf{n}) \right\rangle,
\label{eq:levy_lieb_impurity_functional}
\end{eqnarray}
{and} $\hat{U}_0=U\hat{n}_{0\uparrow}\hat{n}_{0\downarrow}$. The latter
functional is the
analog of the LL functional for a partially-interacting system
where the on-site Coulomb interaction $U$ is switched off on the whole
lattice except on one site labelled as $i=0$. The latter is referred to
as the ``impurity'' whereas the region of all the remaining
(non-interacting) sites is
called ``bath'' in SOET. Note that, if $\mathbf{n}$ is pure-state impurity-interacting
$v$-representable, the minimizing wavefunction $\Psi^{\rm
imp}(\mathbf{n})$ in Eq.~(\ref{eq:levy_lieb_impurity_functional})
is well-defined and it fulfills the following ground-state Schr\"{o}dinger-like
equation: 
\be\label{eq:imp-Schrodinger_eq_dens}
\hat{H}^{\rm imp}(\mathbf{n})\left\vert\Psi^{\rm
imp}(\mathbf{n})\right\rangle=\mathcal{E}^{\rm imp}(\mathbf{n})\left\vert\Psi^{\rm
imp}(\mathbf{n})\right\rangle,
\ee
where the impurity-interacting density-functional Hamiltonian reads 
\begin{eqnarray}
\hat{H}^{\rm imp}(\mathbf{n}) = \hat{T} +
\hat{U}_0+ \sum_{i} v_{i}^{\rm
emb}(\mathbf{n})\,\hat{n}_{i}.
\label{eq:impurity_hamiltonian}
\end{eqnarray}
The unicity (up to a constant) of the potential $\left\{v_{i}^{\rm
emb}(\mathbf{n})\right\}_{0\leq i\leq L-1}$ is guaranteed by the
Hohenberg--Kohn theorem~\cite{hohenberg_inhomogeneous_1964} that we simply apply in this context to
impurity-interacting Hamiltonians.\\
 
The second term on the right-hand side of
Eq.~(\ref{eq:soet_partition}) is the Hxc functional of the bath. It can
be decomposed into Hx and correlation terms as follows,
\be
\overline{E}_{\rm Hxc}^{\rm
bath}(\mathbf{n})=\dfrac{U}{4}\sum_{i\neq 0}n_i^2+\overline{E}_{\rm c}^{\rm
bath}(\mathbf{n}).
\ee
If we
consider 
the KS decomposition of the impurity-interacting LL functional $F^{\rm
imp}(\mathbf{n})=T_{\rm s}(\mathbf{n})+E_{\rm Hxc}^{\rm
imp}(\mathbf{n})$, where $E_{\rm Hxc}^{\rm
imp}(\mathbf{n})=({U}n^2_0/4)+E_{\rm c}^{\rm
imp}(\mathbf{n})$, it comes 
\begin{eqnarray}
\overline{E}^{\rm bath}_{\rm c}(\mathbf{n}) = E_{\rm c}(\mathbf{n})
- E_{\rm c}^{\rm imp}(\mathbf{n})  
,
\label{eq:bath_hxc_energy}
\end{eqnarray}
where we readily see that the complementary bath correlation functional describes
not only the correlation in the bath but also its coupling
with the correlation effects on the impurity.\\
 
By inserting
Eqs.~(\ref{eq:soet_partition}) and
(\ref{eq:levy_lieb_impurity_functional}) into Eq.~(\ref{eq:gs_energy}),
it can be shown~\cite{senjean_site-occupation_2018} that, in
SOET, the KS equations are replaced by the following self-consistent
ground-state many-body wavefunction equation, 
\begin{eqnarray}
& &\left( \hat{T} + 
\hat{U}_0+
%U\hat{n}_{0\uparrow}\hat{n}_{0\downarrow} + 
\sum_{i}
v_{i}^{\rm emb}
\hat{n}_{i}
%\left(\mathbf{n}^{\rm imp}\right)
\right)| \Psi^{\rm imp} \rangle
%\nonumber\\
%& &
= \mathcal{E}^{\rm imp} | \Psi^{\rm imp} \rangle
,
\label{eq:sc_soet_equation}
\end{eqnarray}
where  
\begin{eqnarray}
v_{i}^{\rm emb}
%v_{i}^{\rm emb}(\mathbf{n})
 = v_{i} + \left.
\frac{\partial \overline{E}^{\rm
bath}_{\rm Hxc}(\mathbf{n})}{\partial n_{i}}
\right|_{\mathbf{n}=\mathbf{n}^{\Psi^{\rm imp} }}-\mu
\label{eq:embedding_potential}
\end{eqnarray}
plays
the role of a density-functional embedding potential for the impurity. 
It ensures that the impurity-interacting wavefunction yields the
site-occupation of the physical Hubbard model. Note that, like in DFT,
the latter (auxiliary) wavefunction is not expected to reproduce 
%\manu{I stopped here}
other observables such as the double occupation.
One needs to add the density-functional contributions from the bath [see
Eq.~(\ref{eq:bath_hxc_energy})] to the bare impurity-interacting double occupation
$d^{\rm imp}$
in order to calculate in-principle-exact and physically meaningful properties. In the
particular case of the {\it uniform} Hubbard model, i.e. when $\mathbf{n}^{\Psi^{\rm imp}}\equiv (n,n,\ldots,n)$, the exact double
occupation reads as follows in SOET:~\cite{senjean_site-occupation_2018}   
\begin{eqnarray}
d = d^{\rm imp} + \dfrac{\partial \overline{e}_{\rm c}^{\rm
bath}(\mathbf{n}^{\Psi^{\rm imp}})}{\partial U},
\label{eq:soet_double_occ}
\end{eqnarray}
where
\begin{eqnarray}
d^{\rm imp} = \langle\Psi^{\rm imp} |
\hat{n}_{0\uparrow}\hat{n}_{0\downarrow} | \Psi^{\rm imp} \rangle,
\label{eq:sdimp_exact_soet}
\end{eqnarray}
and
\be\label{eq:ecbath_exps}
\overline{e}_{\rm c}^{\rm bath}(\mathbf{n})&=&e_{\rm
c}({n}_0)-E^{\rm imp}_{\rm c}(\mathbf{n})
\nonumber\\
&=&\overline{E}^{\rm bath}_{\rm c}(\mathbf{n})-\sum_{i\neq0}e_{\rm
c}({n}_i).
\ee 
The latter functional can be seen as a per-site correlation functional for the bath. The exact
physical per-site energy can be expressed as
follows~\cite{senjean_site-occupation_2018,senjean_multiple_2018}:
\begin{eqnarray}
e &=&  t_{\rm s}({n}) + t\frac{\partial e_{\rm
c}(n)}{\partial t} + U d^{\rm imp}
%\nonumber\\
%& &
 + U\frac{\partial \overline{e}_{\rm c}^{\rm
bath}(\mathbf{n}^{\Psi^{\rm imp}})}{\partial U}\,,
\label{eq:exact_persite_energy}
\end{eqnarray}
where $t_{\rm s}(n)$ is the non-interacting per-site kinetic energy
functional [$t_{\rm s}(n)=-4t\sin(\pi n/2)/\pi$ in the 1D 
case]. 

\subsection{Site-occupation--Green's function embedding theory}\label{subsec:soget}

\subsubsection{Density-functional self-energy}

In the following, we propose a complete reformulation and simplication of SOET based on
the Green's function formalism. We would
like the embedding procedure to remain a functional of the density, unlike in conventional approaches like
DMFT or SEET where the Green's function is the basic variable. For that
purpose, we
start from the impurity-interacting many-body
wavefunction in Eq.~(\ref{eq:imp-Schrodinger_eq_dens}) and consider the
corresponding (retarded) equilibrium zero-temperature   
frequency-dependent one-particle Green's function ${\bf
G}^{\rm imp}\left(\mathbf{n},\omega\right)\equiv\left\{G^{\rm
imp}_{i\sigma,j\sigma'}\left(\mathbf{n},\omega\right)\right\}_{
i,j,\sigma,\sigma'}$ with $0\leq
i,j\leq L-1$, which we simply refer to as the
impurity Green's function in the following. Its elements in the Lehmann representation are defined as follows: 
\begin{widetext}
\begin{eqnarray}
G^{\rm imp}_{i\sigma,j\sigma'}\left(\mathbf{n},\omega\right) &=&  
\left
\langle
%\Psi^{\rm imp}(\mathbf{n})
%\middle\vert
\hat{a}_{i\sigma}
\frac{
1} 
{\omega + \mathcal{E}^{\rm imp}(\mathbf{n})- \hat{H}^{\rm imp}(\mathbf{n}) + \im\eta}
\hat{a}_{j\sigma'}^{\dagger}
%\middle\vert
% \Psi^{\rm
%imp}(\mathbf{n})
\right\rangle_{\Psi^{\rm imp}(\mathbf{n})}
%\nonumber\\
%\nonumber\\
%&&
 +
\left\langle
%\Psi^{\rm imp}(\mathbf{n}) \middle\vert
\hat{a}_{j\sigma'}^{\dagger}\frac{
1
}
{\omega - \mathcal{E}^{\rm imp}(\mathbf{n})  + \hat{H}^{\rm imp}(\mathbf{n})  + \im\eta}
 \hat{a}_{i\sigma} 
%\middle\vert \Psi^{\rm
%imp}(\mathbf{n})
\right\rangle_{\Psi^{\rm imp}(\mathbf{n})},
\nonumber\\
\label{eq:impurity_green_function_lehmann}
\end{eqnarray}
\end{widetext}
where $\eta\rightarrow0^+$.
%%%%%%%%%%%%%%%%%%%%%%%%%%%%%%%%%%%%%%%%%%%%%%%%%%%%%%%%%%%%%%%
\iffalse%%%%%%%%%%%%%%%%%%% commented by Manu %%%%%%%%%%%%%%%%%
\begin{eqnarray}
G_{i\sigma,j\sigma'}^{\rm imp}[\mathbf{n}](t-t') =
\langle\Psi_{0}^{\rm imp}[\mathbf{n}] | \lbrace\hat{a}_{i\sigma}^{\dagger}(t),\hat{a}_{j\sigma'}(t') \rbrace | \Psi_{0}^{\rm imp}[\mathbf{n}] \rangle
\nonumber\\
\label{eq:impurity_green_function_time}
\end{eqnarray}
where
\begin{eqnarray}
\hat{a}_{i\sigma}(t) =e^{\im\hat{H}^{\rm imp}[\mathbf{n}] t} \hat{a}_{i\sigma} e^{-\im\hat{H}^{\rm imp}[\mathbf{n}] t}
\label{eq:heisenberg_picture}
\end{eqnarray}
and
By using the time invariance $t\rightarrow t - t'$ and $t'\rightarrow 0$, the Fourier transform of Eq.~\ref{eq:impurity_green_function_time} yields the Lehmann representation of  dependent one-particle impurity Green's function and omit the frequency dependence hereafter for the sake of visibility:
This step is fully justified because, according to the Hohenberg-Kohn theorem, the site occupation uniquely fixes the wavefunction and thus also uniquely fixes the Green's function. Moreover, every information that can be extracted from a wavefunction, such as the site occupation, can also be extracted from a Green's function. 
\fi%%%%%%%%%%%%%%%%%%%%%%
%%%%%%%%%%%%%%%%%%%%%%%%%%%%%%%%
Note that, by construction, the impurity Green's function
reproduces the density profile ${\bf n}$:
\begin{eqnarray}
-\frac{1}{\pi}\sum_{\sigma}\int_{-\infty}^{0}\ddroit\omega\,\mathrm{Im}\,\left[G_{i\sigma,
i\sigma}^{\rm imp}\left(\mathbf{n},\omega\right)\right]
=n_{i},
\label{eq:site_occupation_gimp_exact}
\end{eqnarray}
where we integrate up to zero since the chemical
potential is included into $\hat{H}^{\rm imp}(\mathbf{n})$. Note also
that, by analogy
with the fully-interacting case [see Eq.~(A.1) in Ref.~\cite{Potthoff_EPJB03_G_of_Sigma}], one could make 
the one-to-one correspondence between the impurity-interacting Green's
function and the embedding potential (which is itself an implicit
functional of the ground-state density) 
more explicit by expanding the Green's function through second order in 
$1/\omega$.
\\

Let us now introduce an auxiliary interaction-free Green's function $\mathbfcal{G}^{\rm
imp}\left(\mathbf{n},\omega\right)$ which is obtained by removing from
$\hat{H}^{\rm imp}(\mathbf{n})$ the
interaction on the impurity site. From the point of view of DMFT, 
$\mathbfcal{G}^{\rm
imp}\left(\mathbf{n},\omega\right)$
might be seen as a
density-functional Weiss field, whose explicit expression reads 
\be\label{eq:explicit_exp_Weiss_field}
\mathbfcal{G}^{\rm
imp}\left(\mathbf{n},\omega\right)=\left[\left(\omega+{\rm
i}\eta\right)\mathbf{I}-\mathbf{t}-
\mathbf{v}^{\rm emb}(\mathbf{n})\right]^{-1}
,
\nonumber\\
\ee
where $\mathbf{I}$ is the identity matrix, $\mathbf{t}$ is the matrix
representation of the (one-electron) hopping operator, and $
\mathbf{v}^{\rm emb}(\mathbf{n})
\equiv\left\{\delta_{\sigma\sigma'}\delta_{ij}
v_
i^{\rm emb}
(\mathbf{n})\right\}_{0\leq i,j\leq L-1,\sigma,\sigma'}$ 
is the matrix representation of the local and frequency-independent
embedding potential.
Note that $\mathbfcal{G}^{\rm
imp}\left(\mathbf{n},\omega\right)$ is a functional of the density
$\mathbf{n}$ but it does {\it not} reproduce that density.
\\ 

We can now define a density-functional impurity
self-energy,
\begin{eqnarray}
\mathbf{\Sigma}_{\rm Hxc}^{\rm imp}(\mathbf{n},\omega) = 
\left[\mathbfcal{G}^{\rm
imp}\left(\mathbf{n},\omega\right)\right]^{-1}
- \left[\mathbf{G}^{\rm imp}\left(\mathbf{n},\omega\right)\right]^{-1}
,
\nonumber\\
\label{eq:dyson_equation_impurity_self_energy}
\end{eqnarray}
which is (one of) the central quantity in SOGET.
It is in principle non-local and, when combined with the
impurity Green's function, it gives access to the impurity LL
density-functional energy defined in
Eq.~(\ref{eq:levy_lieb_impurity_functional}). Indeed,
since real algebra can be used to describe the impurity-interacting
density-functional many-body wavefunction $\Psi^{\rm imp}(\mathbf{n})$,
its kinetic energy can be
written as
\be\label{eq:Timp_func}
\langle\Psi^{\rm imp}(\mathbf{n})\vert\hat{T}\vert \Psi^{\rm
imp}(\mathbf{n})\rangle
&=&-t\displaystyle{\sum_{<i,j>,\sigma}} \left\langle\hat{a}_{i\sigma}^{\dagger}\hat{a}_{j\sigma}
 +    \mathrm{h.c.}\right\rangle_{\Psi^{\rm imp}(\mathbf{n})}
\nonumber\\
&=&
-2t\displaystyle{\sum_{<i,j>,\sigma}} \left\langle\hat{a}_{j\sigma}^{\dagger}\hat{a}_{i\sigma}
\right\rangle_{\Psi^{\rm imp}(\mathbf{n})},
\ee 
where, according to
Eq.~(\ref{eq:impurity_green_function_lehmann}),
\be\label{eq:1RDM_from_GF}
\left\langle\hat{a}_{j\sigma}^{\dagger}\hat{a}_{i\sigma}
\right\rangle_{\Psi^{\rm imp}(\mathbf{n})}&=&
-\dfrac{1}{\pi}
\int_{-\infty}^{0}\ddroit\omega\,\mathrm{Im}\,      \left[G_{i\sigma,
j\sigma}^{\rm imp}\left(\mathbf{n},\omega\right)\right]
.
\nonumber\\
\ee
We stress again that the chemical potential is included into
$\hat{H}^{\rm imp}(\mathbf{n})$, hence the integration up to zero in
Eq.~(\ref{eq:1RDM_from_GF}).
In addition, by deriving the equation of motion for the
impurity-interacting Green's function, it can be shown, in complete analogy
with conventional Green's function theory, that the density-functional impurity interaction
energy reads
\be\label{eq:Uimp_func_from_SE-GF}
&&\langle\Psi^{\rm imp}(\mathbf{n})\vert\hat{U}_0\vert \Psi^{\rm
imp}(\mathbf{n})\rangle=U\left\langle
\hat{n}_{0\uparrow}\hat{n}_{0\downarrow}
\right\rangle_{\Psi^{\rm imp}(\mathbf{n})}
\nonumber\\
&&=
-\dfrac{1}{\pi}\int_{-\infty}^{0}\ddroit\omega\,\mathrm{Im}\,      \left[
\mathbf{\Sigma}_{\rm Hxc}^{\rm imp}(\mathbf{n},\omega)\mathbf{G}^{\rm imp}\left(\mathbf{n},           \omega\right)
\right]_{0\sigma',0\sigma'},
\ee  
where $\sigma'$ refers to a spin up or spin down state [for
simplicity, we restrict
the discussion to cases where the Green's function is the same for up and down spins]. 
From Eqs.~(\ref{eq:levy_lieb_impurity_functional}),
(\ref{eq:Timp_func}), (\ref{eq:1RDM_from_GF})
and~(\ref{eq:Uimp_func_from_SE-GF}), we obtain the following expression: 
\be\label{eq:Fimp_GF-Sigma}
&&F^{\rm imp}(\mathbf{n})=\frac{2t}{\pi}\sum_{<i,j>,\sigma}
\int_{-\infty}^{0}\ddroit\omega\,\mathrm{Im}\,      \left[G_{i\sigma,
j\sigma}^{\rm imp}\left(\mathbf{n},\omega\right)\right]
\nonumber\\
&&
-\dfrac{1}{\pi}\int_{-\infty}^{0}\ddroit\omega\,\mathrm{Im}\,      \left[
\mathbf{\Sigma}_{\rm Hxc}^{\rm imp}(\mathbf{n},\omega)\mathbf{G}^{\rm imp}\left(\mathbf{n},\omega\right)
\right]_{0\sigma',0\sigma'}.
\ee   
Combining
Eq.~(\ref{eq:Fimp_GF-Sigma}) with Eq.~(\ref{eq:soet_partition}) leads to
a formally exact SOGET where, like in SOET, the bath is described with a density functional
and, unlike in SOET, a Green's function is used to describe the
impurity-interacting
system (instead of a many-body wavefunction).\\

Interestingly, if we introduce the KS potential into the
density-functional embedding one [see
Eqs.~(\ref{eq:bath_hxc_energy}) and (\ref{eq:embedding_potential})],
\be\label{eq:vemb_decomp_KS-imp}
v_{i}^{\rm
emb}(\mathbf{n})&=&v_i^{\rm KS}(\mathbf{n})-
\frac{\partial {E}_{\rm Hxc}(\mathbf{n})}{\partial n_{i}}
+\frac{\partial \overline{E}^{\rm
bath}_{\rm Hxc}(\mathbf{n})}{\partial n_{i}}
\nonumber\\
&=&v_i^{\rm KS}(\mathbf{n})-\frac{\partial {E}^{\rm imp}_{\rm Hxc}(\mathbf{n})}{\partial n_{i}}
\nonumber\\
&=&v_i^{\rm KS}(\mathbf{n})-v^{\rm imp}_{{\rm Hxc},i}(\mathbf{n}),
\ee      
we deduce from Eqs.~(\ref{eq:explicit_exp_Weiss_field}) and
(\ref{eq:dyson_equation_impurity_self_energy}) the following Dyson
equation, 
\be\label{eq:exact_imp_Sigma_exp}
\mathbf{\Sigma}_{\rm Hxc}^{\rm imp}(\mathbf{n},\omega) &=& 
\left[\mathbf{G}^{\rm KS}\left(\mathbf{n},\omega\right)\right]^{-1}
- \left[\mathbf{G}^{\rm imp}\left(\mathbf{n},\omega\right)\right]^{-1}
\nonumber\\
&&+{\mathbf{v}}_{\rm Hxc}^{\rm imp}(\mathbf{n}),
\ee
or, equivalently,
\be\label{eq:imp_Dyson_eq_from_KS}
&&\mathbf{G}^{\rm imp}\left(\mathbf{n},\omega\right)=\mathbf{G}^{\rm
KS}\left(\mathbf{n},\omega\right)
\nonumber\\
&&+\mathbf{G}^{\rm
KS}\left(\mathbf{n},\omega\right)\left(\mathbf{\Sigma}_{\rm
Hxc}^{\rm imp}(\mathbf{n},\omega)-{\mathbf{v}}_{\rm Hxc}^{\rm
imp}(\mathbf{n})\right)\mathbf{G}^{\rm
imp}\left(\mathbf{n},\omega\right),
\nonumber\\
\ee
where $\mathbf{G}^{\rm KS}\left(\mathbf{n},\omega\right)
=\left[\left(\omega+{\rm
i}\eta\right)\mathbf{I}-\mathbf{t}-
\mathbf{v}^{\rm KS}(\mathbf{n})
\right]^{-1}$ is the
non-interacting KS Green's function. Comparing
Eq.~(\ref{eq:exact_imp_Sigma_exp}) with
Eqs.~(\ref{eq:imp-Schrodinger_eq_dens}),
(\ref{eq:impurity_hamiltonian}), and (\ref{eq:vemb_decomp_KS-imp})
reveals a key difference between SOET
and SOGET. While the former generates the
impurity-interacting many-body wavefunction with density $\mathbf{n}$
from the
local and frequency-independent potential ${\mathbf{v}}_{\rm Hxc}^{\rm
imp}(\mathbf{n})$, SOGET is expected to generate the corresponding
Green's function from the KS one. For that purpose, 
non-local {\it and} frequency-dependent corrections to ${\mathbf{v}}_{\rm Hxc}^{\rm 
imp}(\mathbf{n})$ are in principle needed. These corrections will be
contained in the correlation part of the impurity-interacting self-energy that needs
to be modelled.\\          

Finally, since the KS and impurity-interacting Green's functions reproduce the same density
profile ${\bf n}$, i.e. 
\be\label{eq:dens_imp_equal_densKS}
&&
\sum_{\sigma}\int_{-\infty}^{0}\ddroit\omega\,\mathrm{Im}\,      \left[G_{i\sigma,
i\sigma}^{\rm imp}\left(\mathbf{n},\omega\right)\right]
\nonumber\\
=
&&\sum_{\sigma}\int_{-\infty}^{0}\ddroit\omega\,\mathrm{Im}\,      \left[G_{i\sigma,
i\sigma}^{\rm KS}\left(\mathbf{n},\omega\right)\right],
\ee
we obtain, by inserting Eq.~(\ref{eq:imp_Dyson_eq_from_KS}) into
Eq.~(\ref{eq:dens_imp_equal_densKS}), a Sham--Schl\"{u}ter-like equation
for the impurity-interacting system:
\begin{widetext}
\be\label{eq:imp_SSE}
\sum_{\sigma}\int_{-\infty}^{0}\ddroit\omega\,\mathrm{Im}\,      \left[
\mathbf{G}^{\rm
KS}\left(\mathbf{n},\omega\right)\mathbf{\Sigma}_{\rm
Hxc}^{\rm imp}(\mathbf{n},\omega)
\mathbf{G}^{\rm
imp}\left(\mathbf{n},\omega\right)
\right]_{i\sigma,
i\sigma}
=
\sum_{\sigma,j}{v}_{{\rm Hxc},j}^{\rm
imp}(\mathbf{n})\int_{-\infty}^{0}\ddroit\omega\,\mathrm{Im}\,      \left[
{G}_{i\sigma,j\sigma}^{\rm
KS}\left(\mathbf{n},\omega\right)
{G}_{j\sigma,i\sigma}^{\rm
imp}\left(\mathbf{n},\omega\right)
\right].
\nonumber\\
\ee
\end{widetext}
This relation, which explicitly
connects the local and frequency-independent potential
${\mathbf{v}}_{\rm Hxc}^{\rm imp}(\mathbf{n})$ to the non-local and
frequency-dependent self-energy $\mathbf{\Sigma}_{\rm
Hxc}^{\rm imp}(\mathbf{n},\omega)$, is a stringent condition that could
be used, for example, in the development of 
approximate embedding potentials. This is left for
future work. 
%In the rest of the paper, we show how SOGET can be turned
%into a practical computational method that bypasses the expensive
%calculation of the SOET wavefunction. 
 
\subsubsection{Self-consistency loop in SOGET}
We explain in this section how the impurity-interacting Green's function
can be determined self-consistently from the density-functional impurity
self-energy. Our starting point will be the SOET
Eq.~(\ref{eq:sc_soet_equation}) where the quantity to be determined
self-consistently is the impurity-interacting many-body wavefunction. We
want to bypass this step, which is computationally demanding, and
propose an alternative equation where the unknown quantity is the
corresponding Green's function.\\
 
Let us assume that we have at hand both the correlation functional for
the bath {\it and} the impurity density-functional self-energy (these
two
quantities will of course be approximated later on). We denote
$\mathbf{G}^{\rm imp}\left(\omega\right)$ the Green's function
constructed from the solution $\Psi^{\rm imp}$ to the self-consistent SOET
Eq.~(\ref{eq:sc_soet_equation}),
%\begin{widetext}
\begin{eqnarray}
&&G^{\rm imp}_{i\sigma,j\sigma'}\left(\omega\right) =  
\left
\langle
%\Psi^{\rm imp}(\mathbf{n})
%\middle\vert
\hat{a}_{i\sigma}
\frac{
1} 
{\omega + \mathcal{E}^{\rm imp}- \hat{H}^{\rm imp} + \im\eta}
\hat{a}_{j\sigma'}^{\dagger}
%\middle\vert
% \Psi^{\rm
%imp}(\mathbf{n})
\right\rangle_{\Psi^{\rm imp}}
\nonumber\\
&&
+
\left\langle
%\Psi^{\rm imp}(\mathbf{n}) \middle\vert
\hat{a}_{j\sigma'}^{\dagger}\frac{
1
}
{\omega - \mathcal{E}^{\rm imp}  + \hat{H}^{\rm imp}  + \im\eta}
 \hat{a}_{i\sigma} 
%\middle\vert \Psi^{\rm
%imp}(\mathbf{n})
\right\rangle_{\Psi^{\rm imp}},
\label{eq:sc-impurity_green_function_lehmann}
\end{eqnarray}
%\end{widetext}
where 
\be
&&\hat{H}^{\rm imp}= \hat{T} + 
\hat{U}_0
\nonumber\\
&&
+\sum_{i}
\left[v_{i}-\mu+
\left.
\frac{\partial \overline{E}^{\rm
bath}_{\rm Hxc}(\mathbf{n})}{\partial n_{i}}
\right|_{\mathbf{n}=\mathbf{n}^{\Psi^{\rm imp} }}
\right]
\hat{n}_{i}.
\ee 
By construction, the density profile ${\mathbf{n}}^{\mathbf{G}^{\rm
imp}}\equiv\left\{n_i^{\mathbf{G}^{\rm
imp}}\right\}_{0\leq i\leq L-1}$ generated from $\mathbf{G}^{\rm
imp}\left(\omega\right)$ equals the density profile of $\Psi^{\rm imp}$: 
\be
n_i^{\mathbf{G}^{\rm
imp}}&=&-\dfrac{1}{\pi}\sum_\sigma
\int_{-\infty}^{0}\ddroit\omega\,\mathrm{Im}\,      \left[G_{i\sigma,
i\sigma}^{\rm imp}\left(\omega\right)\right]
\nonumber\\
&=&\sum_\sigma
\bra{\Psi^{\rm imp}}\hat{a}_{i\sigma}^{\dagger}\hat{a}_{i\sigma}\ket{\Psi^{\rm imp}}
\nonumber\\
&=&n_i^{\Psi^{\rm imp}},
\ee
which is itself equal to the density of the physical (Hubbard) system if no
approximation is made.\\

Since, as readily seen from the SOET
Eq.~(\ref{eq:sc_soet_equation}), $\Psi^{\rm imp}$ is the ground state of
an impurity-interacting system with density $\mathbf{n}^{\Psi^{\rm imp}
}={\mathbf{n}}^{\mathbf{G}^{\rm
imp}}$ and embedding potential
\be\label{eq:vemb_nGimp}
\mathbf{v}^{\rm emb}(\mathbf{n}^{\mathbf{G}^{\rm
imp}})=\mathbf{v}-\mu+\left.\frac{\partial \overline{E}^{\rm
bath}_{\rm Hxc}(\mathbf{n})}{\partial \mathbf{n}}\right|_{\mathbf{n}=\mathbf{n}^{\mathbf{G}^{\rm
imp}}},
\ee    
we conclude [see Eq.~(\ref{eq:impurity_hamiltonian})] that
\be
\hat{H}^{\rm imp}\left(\mathbf{n}
^{\mathbf{G}^{\rm
imp}}
\right)=\hat{H}^{\rm imp},
\ee
thus leading to
\be
\Psi^{\rm imp}\left(\mathbf{n}
^{\mathbf{G}^{\rm
imp}}
\right)&=&\Psi^{\rm imp},
\nonumber\\
\mathcal{E}^{\rm imp}\left(\mathbf{n}
^{\mathbf{G}^{\rm
imp}}
\right)&=&\mathcal{E}^{\rm imp},
\ee 
so that [see Eqs.~(\ref{eq:impurity_green_function_lehmann}) and
(\ref{eq:sc-impurity_green_function_lehmann})],
\be
\mathbf{G}^{\rm imp}\left({\mathbf{n}}^{\mathbf{G}^{\rm
imp}},\omega\right)=\mathbf{G}^{\rm imp}\left(\omega\right).
\ee
Finally, the latter equation can be rewritten as [see Eq.~(\ref{eq:dyson_equation_impurity_self_energy})]
\be
\left[\mathbf{G}^{\rm imp}\left(\omega\right)\right]^{-1}&=&
\left[\mathbf{G}^{\rm imp}\left({\mathbf{n}}^{\mathbf{G}^{\rm
imp}},\omega\right)\right]^{-1}
\nonumber\\
&=&\left[\mathbfcal{G}^{\rm
imp}\left(\mathbf{n}^{\mathbf{G}^{\rm
imp}},\omega\right)\right]^{-1}
-\mathbf{\Sigma}_{\rm Hxc}^{\rm imp}\left(\mathbf{n}^{\mathbf{G}^{\rm
imp}},\omega\right),
\nonumber\\
\ee
or, equivalently [see Eqs.~(\ref{eq:explicit_exp_Weiss_field}) and (\ref{eq:vemb_nGimp})], 
\be
\left[\mathbf{G}^{\rm imp}\left(\omega\right)\right]^{-1}&=&
\left[\mathbfcal{G}_{\mathbf{v}}^{\rm
imp}\left(\mathbf{n}^{\mathbf{G}^{\rm
imp}},\omega\right)
\right]^{-1}-\mathbf{\Sigma}_{\rm Hxc}^{\rm imp}\left(\mathbf{n}^{\mathbf{G}^{\rm
imp}},\omega\right),
\nonumber\\
\ee
where the density-functional interaction-free Green's function $\mathbfcal{G}_{\mathbf{v}}^{\rm
imp}\left(\mathbf{n},\omega\right)$ is determined from the physical
external potential $\mathbf{v}$ as follows, 
\be
\mathbfcal{G}_{\mathbf{v}}^{\rm
imp}\left(\mathbf{n},\omega\right)=
\left[\left(\omega+\mu+{\rm
i}\eta\right)\mathbf{I}-\mathbf{t}-\mathbf{v}-\frac{\partial \overline{E}^{\rm
bath}_{\rm Hxc}(\mathbf{n})}{\partial \mathbf{n}}\right]^{-1}.
\nonumber\\
\label{eq:v-depdt-weiss_field_green_function}
\ee
Thus we conclude that the impurity-interacting Green's function
$\mathbf{G}^{\rm imp}\left(\omega\right)$ fulfills  
the following self-consistent SOGET
equation, 
\be\label{eq:sc-soget_eq_Sigma}
\mathbf{\Sigma}_{\rm Hxc}^{\rm imp}\left(\mathbf{n}^{\mathbf{G}},\omega\right) = 
\left[\mathbfcal{G}_{\mathbf{v}}^{\rm
imp}\left(\mathbf{n}^{\mathbf{G}},\omega\right)\right]^{-1}
- \left[\mathbf{G}\left(\omega\right)\right]^{-1},
%\nonumber\\
\ee
or, equivalently,
\be
\mathbf{G}\left(\omega\right)=\left[\left[\mathbfcal{G}_{\mathbf{v}}^{\rm
imp}\left(\mathbf{n}^{\mathbf{G}},\omega\right)\right]^{-1}-\mathbf{\Sigma}_{\rm
Hxc}^{\rm imp}\left(\mathbf{n}^{\mathbf{G}},\omega\right)\right]^{-1},
\label{eq:sc_soget_dyson_equation}
\ee
which can be seen as an in-principle-exact density-functional version of the self-consistency loop in
DMFT. 
The lighter notation $\mathbf{G}$ (without the superscript
``imp'') is used to make the self-consistent character of
Eq.~(\ref{eq:sc_soget_dyson_equation}) more visible. 
\iffalse%%%%
Note that it might
be possible to recover Eq.~(\ref{eq:sc_soget_dyson_equation}) directly from a
variational principle by introducing the impurity-interacting analog of
the Luttinger--Ward functional. The latter would become a functional of
the density through the Green's function that is itself functional of
the density.
\fi%%%%     
\\

Turning to uniform systems ($\mathbf{v}\equiv0$), the self-consistently converged solution to the SOGET
Eq.~(\ref{eq:sc_soget_dyson_equation}) can be combined with the SOET
expressions for the double occupation and per-site energy [see
Eqs.~(\ref{eq:soet_double_occ}), (\ref{eq:exact_persite_energy}), and
(\ref{eq:Fimp_GF-Sigma})], thus leading
to the final SOGET expressions,  
\begin{eqnarray}
d &=&-\dfrac{1}{\pi U}\int_{-\infty}^{0}\ddroit\omega\,\mathrm{Im}\,      \left[
\mathbf{\Sigma}_{\rm Hxc}^{\rm imp}\left(\mathbf{n}^{\mathbf{G}^{\rm
imp}},\omega\right)\mathbf{G}^{\rm imp}\left(\omega\right)
\right]_{0\sigma',0\sigma'} 
\nonumber\\
&&+ \dfrac{\partial \overline{e}_{\rm c}^{\rm
bath}\left(\mathbf{n}^{\mathbf{G}^{\rm imp}}\right)}{\partial U},
\nonumber\\
\label{eq:dimp_exact_soget}
\end{eqnarray}
and
\be\label{eq:exact_ener_er_site_exp_GF}
e &=&  \left[t_{\rm s}({n}) + t\frac{\partial e_{\rm
c}(n)}{\partial t}\right]_{n={n}_0^{\mathbf{G}^{\rm imp}}} +Ud,
\ee
where ${n}_0^{\mathbf{G}^{\rm imp}}$ is the impurity site occupation.\\

%Let us stress that, so far, no approximation has been made. 
%Unlike
%DMFT, SOGET is
%formally exact in any dimension
In addition to the non-locality of the
impurity-interacting self-energy (that will be neglected in the rest of this work),
the use of a correlation density functional for describing the bath, which is
inherited from SOET, plays a crucial role in making SOGET in principle
exact, whatever the dimension of the system is. This is an important
difference with DMFT, which is only exact in the infinite dimension
limit~\cite{metzner_correlated_1989}. As illustrated in Sec.~\ref{sec:res_disc}, SOGET can actually 
describe one-dimensional systems accurately. 
 
\subsection{Opening of the gap and derivative
discontinuities}\label{subsec:gap_DD_spec_fun}

We discuss in this section the calculation of the chemical potential $\mu\equiv\mu(n)$ as a
function of the filling [from now on SOGET is applied to the 1D uniform
Hubbard model, i.e. $\mathbf{v}\equiv0$]. Like in KS-DFT, the
impurity-interacting Green's function of SOGET is expected to
reproduce the physical density [the correct filling in this case], {\it
not} the physical spectral function. Nevertheless, as the former is
determined from the latter, we need to explain how the opening of gaps
can effectively occur in SOGET. This is a crucial point when it comes to
model density-driven Mott--Hubbard transitions (see Sec.~\ref{subsec:MH-transition}). For clarity, we first address this problem in
both physical and KS systems, thus highlighting the importance
of derivative discontinuities in DFT-based methods. 
%In the following,   
%$\mathbf{G}(\omega)$ will refer successively to the physical (fully-interacting), KS
%(non-interacting), and impurity-interacting Green's functions.  

\subsubsection{Physical system}\label{subsubsec:chem_pot_physical}

In conventional Green's function theory, the Green's function of the
physical fully-interacting system fulfills the following self-consistent
Dyson equation,
\be\label{eq:SC-physical_Dysoneq}
\left[\mathbf{G}\left(\omega\right)\right]^{-1}&=&\left(\omega+\mu+{\rm 
i}\eta\right)\mathbf{I}-\mathbf{t}
%-\left(\dfrac{U}{2}n_0^{\mathbf{G}}\right)\mathbf{I}
%\nonumber\\
%&&
-\mathbf{\Sigma}_{\rm Hxc}\left(\mathbf{G},\omega\right),
\ee
where, unlike in SOGET, the self-energy is a functional of the
Green's function, not the density.\\

When crossing half-filling (i.e. $n=1$), the on-site two-electron
repulsion, which is fully described by the self-energy, induces an
opening of the energy gap~\cite{lima_density-functional_2002}. In order
to guarantee that the density continuously vary from $1^-$ to $1^+$,
i.e.  
\be\label{eq:continuity_cond_dens}
&&-\dfrac{1}{\pi}\sum_\sigma
\int_{-\infty}^{0}\ddroit\omega\,\mathrm{Im}\,      \left.\left[G_{0\sigma,
0\sigma}\left(\omega\right)\right]\right|_{n\rightarrow 1^-}
\nonumber\\
&&=
-\dfrac{1}{\pi}\sum_\sigma
\int_{-\infty}^{0}\ddroit\omega\,\mathrm{Im}\,      \left.\left[G_{0\sigma,
0\sigma}\left(\omega\right)\right]\right|_{n\rightarrow 1^+}
\nonumber\\
&&=1,
\ee
even though the self-energy induces a gap opening, the chemical
potential has to exhibit a discontinuity at half-filling: 
\be\label{eq:DD_phys_chem_pot}
\mu_-=\left.\mu(n)\right|_{n\rightarrow 1^-}
\neq 
\mu_+=\left.\mu(n)\right|_{n\rightarrow 1^+}.
\ee
Note that, since
\be
\mu(n)=\partial e(n)/\partial n\equiv\partial E(N)/\partial N,
\ee 
the difference $\mu_+-\mu_-$ in chemical potential corresponds to the
physical fundamental gap 
\be\label{eq:fund_gap_exp}
E_g=E(L-1)+E(L+1)-2E(L)
\ee 
of the half-filled ($N=L$) Hubbard system. 

\subsubsection{Kohn--Sham system}

Let us now turn to the KS equation which can be
rewritten in terms of the Green's function as follows: 
\be\label{eq:sc-GF_eq_KS}
\left[\mathbf{G}\left(\omega\right)\right]^{-1}&=&\left(\omega+\mu+{\rm 
i}\eta\right)\mathbf{I}-\mathbf{t}
\nonumber\\
&&-\left[\dfrac{U}{2}n_0^{\mathbf{G}}+\left.\dfrac{\partial e_{\rm
c}(n)}{\partial n}\right|_{n=n_0^{\mathbf{G}}}\right]\mathbf{I}
,
\ee
where $\mu$ is the {\it physical} chemical potential that also appears
in Eq.~(\ref{eq:SC-physical_Dysoneq}). The self-consistently converged solution is the non-interacting KS
Green's function $\mathbf{G}^{\rm KS}(\omega)$. If we assume that the
correlation potential is discontinuous at half-filling [it will become
clear in the following that it can not be otherwise], the KS Green's
functions in the left and right half-filled limits can be connected as
follows, according to Eq.~(\ref{eq:DD_phys_chem_pot}):   
\be\label{eq:KS_GF_cDD}
\left.\left[\mathbf{G}^{\rm KS}\left(\omega\right)\right]^{-1}\right|_{n\rightarrow 1^+}
&=&\left.\left[\mathbf{G}^{\rm KS}\left(\omega+\mu_+-\mu_--\Delta_{\rm c}\right)\right]^{-1}\right|_{n\rightarrow
1^-}, 
\nonumber\\
\ee
where
\be
\Delta_{\rm c}=
\left.\dfrac{\partial e_{\rm
c}(n)}{\partial n}\right|_{n=1^+}
-
\left.\dfrac{\partial e_{\rm
c}(n)}{\partial n}\right|_{n=1^-}
\ee
is the (correlation) derivative discontinuity. Note that $\mathbf{G}^{\rm
KS}(\omega)$ is expected to reproduce the exact density. Therefore it should fulfill the
continuity condition in Eq.~(\ref{eq:continuity_cond_dens}). Combining
the latter condition with the fact that $\mathbf{G}^{\rm
KS}(\omega)$
is a {\it non-interacting} Green's function leads to the following
relation, 
\be\label{eq:KS_GF_gap_orb}
\left.\left[\mathbf{G}^{\rm KS}\left(\omega\right)\right]^{-1}\right|_{n\rightarrow 1^+}
&=&\left.\left[\mathbf{G}^{\rm KS}\left(\omega+\mu^{\rm KS}_+-\mu^{\rm
KS}_-\right)\right]^{-1}\right|_{n\rightarrow
1^-}, 
\nonumber\\
\ee
where $\mu^{\rm KS}_+-\mu^{\rm
KS}_-$ is the KS orbital gap. Thus we recover from Eqs.~(\ref{eq:KS_GF_cDD}) and
(\ref{eq:KS_GF_gap_orb}) the well-known DFT expression for the fundamental
gap~\cite{perdew1983physical} (without the exchange derivative discontinuity
correction
as we work with the Hubbard model),  
\be
\mu_+-\mu_-&=&\mu^{\rm KS}_+-\mu^{\rm
KS}_-+\Delta_{\rm c}.
\ee
In the thermodynamic limit of the 1D Hubbard model, the KS gap becomes
zero ($\mu^{\rm KS}_+=\mu^{\rm
KS}_-=0$) so that the opening of the physical gap fully relies on the derivative
discontinuity in the correlation potential:
\be
\mu_+-\mu_-&\underset{L\rightarrow+\infty}{\longrightarrow}&\Delta_{\rm c}.
\ee
Interestingly, the
BALDA functional incorporates this
feature~\cite{lima_density-functional_2002}, unlike conventional {\it ab
initio} functionals.
Let us stress that the physical gap opening will {\it not} appear explicitly in the KS spectral
function. Indeed, as readily seen from Eq.~(\ref{eq:KS_GF_gap_orb}), the
KS Green's function only describes a gapless non-interacting
system:
\be
\left.\left[\mathbf{G}^{\rm KS}\left(\omega\right)\right]^{-1}\right|_{n\rightarrow 1^+}
&\underset{L\rightarrow+\infty}{\longrightarrow}&\left.\left[\mathbf{G}^{\rm KS}\left(\omega\right)\right]^{-1}\right|_{n\rightarrow 1^-}
\nonumber\\
&=&
\left(\omega+{\rm
i}\eta\right)\mathbf{I}-\mathbf{t},
\ee
where we used the relation $\mu^{\rm
KS}_-\equiv\mu_--(U/2)-\left.\partial e_{\rm c}(n)/\partial
n\right|_{n=1^-}=0$ [see Eq.~(\ref{eq:sc-GF_eq_KS})]. The {\it ad hoc}
derivative discontinuity correction to the KS gap is the key ingredient for
describing (effectively) the opening of the physical gap.  

\subsubsection{Impurity-interacting
system}\label{subsubsec:spec_fun_imp_int_system}

We finally turn to the SOGET Eq.~(\ref{eq:sc-soget_eq_Sigma}) that we
propose to rewrite as follows for analysis purposes:  
\be\label{eq:exact_sc_eq_imp-bath}
&&\left[\mathbf{G}\left(\omega\right)\right]^{-1}=\left(\omega+\mu+{\rm
i}\eta\right)\mathbf{I}-\mathbf{t}
\nonumber\\
&&-\mathbf{I}^{\rm bath}\left.
\left[\dfrac{U}{2}n_0+\dfrac{\partial e_{\rm
c}(n_0)}{\partial n_0}
-
\dfrac{\partial E^{\rm imp}_{\rm c}({\mathbf{n}})}{\partial \mathbf{n}}
\right]
\right|_{\mathbf{n}=\mathbf{n}^{\mathbf{G}}}
\nonumber\\
&&+\mathbf{I}^{\rm imp}
\left[
\dfrac{\partial E^{\rm imp}_{\rm c}({\mathbf{n}})}{\partial n_0}
-\left.\dfrac{\partial e_{\rm
c}(n_0)}{\partial n_0}
\right]
\right|_{\mathbf{n}=\mathbf{n}^{\mathbf{G}}}
%\times
-\mathbf{\Sigma}_{\rm
Hxc}^{\rm imp}\left(\mathbf{n}^{\mathbf{G}},\omega\right)
,
\nonumber\\
\ee
where $\mathbf{I}^{\rm imp}$ and $\mathbf{I}^{\rm bath}$ are the
projectors onto the impurity and bath orbital spaces, respectively
($\mathbf{I}=\mathbf{I}^{\rm imp}+\mathbf{I}^{\rm bath}$). The
self-consistently converged solution to the latter equation is the impurity-interacting Green's
function $\mathbf{G}^{\rm imp}(\omega)$.\\

As readily seen from Eq.~(\ref{eq:exact_sc_eq_imp-bath}), the inverse of
the Green's function is determined, within the impurity orbital space,
from two quantities: a (frequency-independent) density-functional
potential and a frequency-dependent self-energy [fourth and fifth terms
on the right-hand side of the equation]. The former is not expected to
exhibit a derivative discontinuity at half-filling, even though both
impurity-interacting and fully-interacting correlation potentials do.
This statement is based on the fact that, in the
atomic limit, the derivative
discontinuities do cancel each
other~\cite{senjean_site-occupation_2018}. As a result, in SOGET, we expect the
impurity-interacting self-energy to be responsible for the shift $\mu_-\rightarrow \mu_+$ in 
chemical potential on the {\it impurity} site. In the {\it bath},  
this shift is induced by the derivative discontinuity in the
fully-interacting 
correlation potential $\partial e_{\rm c }(n)/\partial n$ [see the third
term on the right-hand side of Eq.~(\ref{eq:exact_sc_eq_imp-bath})].
Note that we do not expect the impurity-interacting correlation
potential $\partial {E}_{\rm c}^{\rm imp}(\mathbf{n})/\partial n_i$ to exhibit
a derivative discontinuity in the bath (i.e. for $i>0$). Again, this statement relies on
what can be seen in the atomic limit [see Eq.~(D15) in
Ref.~\cite{senjean_site-occupation_2018}].\\

In conclusion, in order to
effectively open the physical gap (i.e. shift the chemical potential on
{\it all} sites), the presence of derivative
discontinuities in the bath seems to be essential. For that reason, like
in KS-DFT, we do not expect 
the impurity-interacting spectral function of SOGET to exhibit a gap
opening. This is actually confirmed
by the numerical calculations presented in Sec.~
\ref{subsec:spectr_func}.

\iffalse%%%%%%%%%%%%%%% with Sigma_c %%%
\be
&&\left[\mathbf{G}\left(\omega\right)\right]^{-1}=\left(\omega+\mu+{\rm
i}\eta\right)\mathbf{I}-\mathbf{t}
\nonumber\\
&&-
\left[\dfrac{U}{2}n_0^{\mathbf{G}}+\left.\dfrac{\partial e_{\rm
c}(n)}{\partial n}\right|_{n=n_0^{\mathbf{G}}}\right]\mathbf{I}^{\rm bath}
\nonumber\\
&&-
\left[\dfrac{U}{2}n_0^{\mathbf{G}}+\left.\dfrac{\partial e_{\rm
c}(n)}{\partial n}\right|_{n=n_0^{\mathbf{G}}}\right]\mathbf{I}^{\rm imp}
-\mathbf{\Sigma}_{\rm
c}^{\rm imp}\left(\mathbf{n}^{\mathbf{G}},\omega\right)
\ee
where
\be
\mathbf{\Sigma}_{\rm
c}^{\rm imp}\left(\mathbf{n},\omega\right)=
\mathbf{\Sigma}_{\rm
Hxc}^{\rm imp}\left(\mathbf{n},\omega\right)-\frac{\partial {E}^{\rm
imp}_{\rm Hxc}(\mathbf{n})}{\partial \mathbf{n}}
\ee
\fi%%%%%%%%%%%
 
\section{Approximations}\label{sec:approximations}

In order to turn SOGET into a practical computational method we first need
a density-functional approximation for the bath [i.e. $\overline{E}^{\rm
bath}_{\rm c}(\mathbf{n})$ or, equivalently, $e_{\rm
c}({n})$ and $E^{\rm imp}_{\rm c}(\mathbf{n})$], like in SOET.
The approximations which are used in this work are
briefly reviewed in Sec.~\ref{subsec:approx_soet}. The new ingredient to
be modelled is the impurity-interacting density-functional
self-energy $\mathbf{\Sigma}_{\rm Hxc}^{\rm imp}(\mathbf{n},\omega)$ for
which a local approximation based on the Anderson dimer is constructed in
Secs.~\ref{subsec:local-Sigma} and \ref{subsec:approx_sfimp}. 

\subsection{Approximations to density-functional correlation energies}\label{subsec:approx_soet}

In the particular case of the 1D Hubbard model, the per-site correlation energy functional can be
described within BALDA~\cite{lima_density_2003}:
\begin{eqnarray}
e_{\rm c}(n) \rightarrow e_{\rm c}^{\rm BA}(n) \, .
\label{eq:ec_to_ec_balda}
\end{eqnarray}
By construction, the BALDA is exact (in the thermodynamic limit) at
half-filling for any $U/t$ value, and for any fillings when $U/t=0$ or
$U/t\rightarrow+\infty$. Following
Ref.~\cite{senjean_site-occupation_2018}, we will assume that the
impurity correlation energy does not vary with the occupations in the
bath, which is an approximation~\cite{senjean_local_2017}:  
\begin{eqnarray}
E_{\rm c}^{\rm imp}(\mathbf{n}) \rightarrow E_{\rm c}^{\rm imp}(n_{0}) \, ,
\label{eq:ecimp_neglect_of_nbath}
\end{eqnarray}
thus leading to the following simplifications in the per-site bath
correlation functional [see Eq.~(\ref{eq:ecbath_exps})], 
\begin{eqnarray}
 \overline{e}_{\rm c}^{\rm bath}(\mathbf{n}) 
 \rightarrow e_{\rm c}^{\rm BA}(n_{0}) - E^{\rm imp}_{\rm c}(n_{0}),
\label{eq:ecbath_approx}
\end{eqnarray}
and in the embedding potential to be used in the interaction-free Green's function [see
Eq.~(\ref{eq:v-depdt-weiss_field_green_function})]:
%\be
%\frac{\partial \overline{E}^{\rm
%bath}_{\rm Hxc}(\mathbf{n})}{\partial n_{i}}&=&(1 -
%\delta_{0i})\left[\dfrac{U}{2}n_{i}
%+\left.\dfrac{\partial e_{\rm c}(n)}{\partial n}\right|_{n=n_i}\right]
%\nonumber\\
%&&+\frac{\partial \overline{e}^{\rm
%bath}_{\rm c}(\mathbf{n})}{\partial n_{i}} 
%\ee
\begin{eqnarray}
%v_{i}^{\rm emb}(n_{0}) 
\frac{\partial \overline{E}^{\rm
bath}_{\rm Hxc}(\mathbf{n})}{\partial n_{i}}
&\rightarrow&  (1 - \delta_{0i})\dfrac{U}{2}n_{0}  +
\left.\dfrac{\partial e_{\rm c}^{\rm BA}(n)}{\partial n}\right|_{n=n_0} 
\nonumber\\
&&- \delta_{i0} \dfrac{\partial E^{\rm imp}_{\rm c}(n_{0})}{\partial
n_{0}}. 
\nonumber\\
\label{eq:approx_embedding_potential}
\end{eqnarray}
Note that, in Eq.~(\ref{eq:approx_embedding_potential}), we assumed that the density profile is uniform (as it
should).\\

Various local density-functional approximations to the impurity
correlation energy have been explored in
Refs.~\cite{senjean_site-occupation_2018} and
\cite{senjean_multiple_2018}. In this work, we will use the one
extracted from the two-level (2L) {\it Anderson}
model~\cite{senjean_local_2017} which can be expressed as follows: 
\begin{eqnarray}
E^{\rm imp}_{\rm c}(n_{0}) \rightarrow E^{\rm imp, 2L}_{\rm
c}(U,n_{0})=E^{\rm 2L}_{\rm c}(U/2,n_{0}),
\label{eq:two_level_impurity_corr_energy}
\end{eqnarray}
where $E^{\rm 2L}_{\rm c}(U,n_{0})$ is the density-functional
correlation energy of the two-electron {\it Hubbard} dimer with on-site
interaction strength $U$. In practice, we use the accurate parameterization of
Carrascal~\etal~\cite{carrascal_hubbard_2015,carrascal_corrigendum:_2016} for computing $E^{\rm
2L}_{\rm c}(U/2,n_{0})$ and its derivatives. Note that the same model
will be considered in Sec.~\ref{subsec:approx_sfimp} in order to
construct an approximate impurity-interacting
self-energy. The combination of the 2L approximation with BALDA in
Eqs.~(\ref{eq:ecbath_approx}) and (\ref{eq:approx_embedding_potential})
will simply be referred to as
2L-BALDA in the following.  

\subsection{Local self-energy approximation}\label{subsec:local-Sigma}

By analogy with DMFT, we make the assumption that the impurity-interacting self-energy introduced in
Eq.~(\ref{eq:dyson_equation_impurity_self_energy}) is local: 
\be\label{eq:local_approx_SOGET}
\mathbf{\Sigma}_{\rm Hxc}^{\rm imp}(\mathbf{n},\omega)\rightarrow
\delta_{\sigma\sigma'}\delta_{i0}\delta_{j0}\,{\Sigma}_{\rm Hxc}^{\rm
imp}({n}_0,\omega). 
\ee
Consequently, the SOGET Eq.~(\ref{eq:sc_soget_dyson_equation}) can be simplified as follows:
\be
{G}\left(\omega\right)=\dfrac{1}{\dfrac{1}{\mathcal{G}^{\rm
imp}\left({n}_0^{{G}},\omega\right)}-{\Sigma}_{\rm Hxc}^{\rm
imp}({n}^G_0,\omega)},
%\Sigma_{\rm Hxc}^{\rm imp}[n_{0},\omega] = G_{00}^{\rm imp,0}[n_{0}]^{-1} - G_{00}^{\rm imp}[n_{0}]^{-1}
\label{eq:local_impurity_self_energy_0}
\ee
where
\be\label{eq:density_from_converged_G}
{n}^G_0=-\dfrac{2}{\pi}\int_{-\infty}^{0}\ddroit\omega\,\mathrm{Im}\,
\left[G\left(\omega\right)\right].
\ee 
As mentioned previously, we assume that spin up and spin down 
Green's functions are equal [hence the factor 2 in
Eq.~(\ref{eq:density_from_converged_G})]. The self-consistently
converged solution to Eq.~(\ref{eq:local_impurity_self_energy_0}) will be an approximation
to ${G}_{0\sigma,0\sigma}^{\rm imp}\left(\omega\right)$. The
approximate interaction-free Green's function on the impurity site  
$\mathcal{G}^{\rm
imp}\left({n}_0,\omega\right)$ can be seen as a
density-functional Weiss field whose final expression reads (see
Appendix~\ref{app:hybridization_function}),
\begin{eqnarray}
\mathcal{G}^{\rm imp}(n_{0},\omega) = \dfrac{1}{\omega + \im\eta + \mu -
v_{0}^{\rm emb}(n_{0}) - \Delta(n_{0},\omega)},
\nonumber\\
\label{eq:local_weiss_field_green_function}
\end{eqnarray}
where  
\be
v^{\rm emb}_0(n_0)&=&\dfrac{\partial e_{\rm c}^{\rm BA}(n_0)}{\partial
n_0}-\dfrac{\partial E^{\rm imp}_{\rm c}(n_{0})}{\partial
n_{0}}, 
\ee
and 
\begin{eqnarray}
\Delta(n_{0},\omega) = \sum_{k}\frac{|V_{0k}|^{2}}{\omega + \im\eta + \mu -
\varepsilon_{k}(n_{0})}
\label{eq:hybridization_function}
\end{eqnarray}
is the analog of the hybridization function in DMFT~\cite{georges_beauty_2016}. The bath orbital energies
$\varepsilon_{k}(n_{0})$ and impurity-bath coupling terms $V_{0k}$  are obtained
by diagonalizing the projection onto the bath of the interaction-free SOET Hamiltonian (further details are given in
Appendix~\ref{appendix:diag_bath}).\\ 

If we use the following exact expression for the chemical potential,
\be
\mu=\mu^{\rm KS}(n_0)+\dfrac{U}{2}n_0+\dfrac{\partial e_{\rm
c}(n_0)}{\partial n_0},
\ee
where the KS chemical potential reads (in 1D)
\be
\mu^{\rm KS}(n_0)=\dfrac{\partial t_{\rm s}(n_0)}{\partial
n_0}=-2t\cos\left(\frac{\pi}{2}n_0\right),
\label{eq:chem_pot_0}
\ee
we see from
Eq.~(\ref{eq:local_weiss_field_green_function}) that, within the
2L-BALDA approximation, the total potential on
the impurity site will be simplified as follows: 
\be\label{eq:2L-BALDA-mu_minus_vemb}
-\mu +
v_{0}^{\rm emb}(n_{0})
&\rightarrow&-\dfrac{U}{2}n_0-\frac{\partial
E_{\rm c}^{\rm imp,2L}(U,n_{0})}{\partial n_{0}}
\nonumber\\
&&-\dfrac{\partial t_{\rm s}(n_0)}{\partial
n_0},
\ee 
while giving in the bath (see
Eq.~(\ref{eq:approx_embedding_potential})),
\be\label{eq:2L-BALDA-mu_minus_vemb_bath}
-\mu +
v_{i}^{\rm emb}(n_{0})
&\xrightarrow{i>0}&
-\dfrac{\partial t_{\rm s}(n_0)}{\partial
n_0}.
\ee 
With the latter simplification, the following substitution can therefore be made in the hybridization
function (see Eq.~(\ref{eq:epsilon_k_n0})):
\be\label{eq:simplification_inside_hyb_func}
\mu -
\varepsilon_{k}(n_{0})&\rightarrow&\mu^{\rm KS}(n_0)+2t\cos(k),
\ee
thus showing that the bath is basically treated within KS DFT.    
Note also that the 2L impurity-interacting correlation potential does not exhibit a derivative discontinuity
at $n_0=1$ for
finite $U/t$ values~\cite{senjean_local_2017}. According to
Eqs.~(\ref{eq:2L-BALDA-mu_minus_vemb}) and (\ref{eq:2L-BALDA-mu_minus_vemb_bath}), in the
half-filled left or right limits
($n_0\rightarrow 1^\mp$), the total potential will therefore be equal to $-U/2$ on the
impurity and it will vanish in the bath, which is exact for half-filled
finite systems~\cite{senjean_site-occupation_2018}. 

\subsection{Two-level density-functional self-energy approximation}\label{subsec:approx_sfimp}

A simple but non-trivial way to design a local density-functional
approximation to the impurity-interacting self-energy consists in applying SOGET to
the {\it two-electron} Hubbard dimer. This idea originates from 
the two-site version of
DMFT~\cite{lange_renormalized_1998,bulla_linearized_2000,potthoff_two-site_2001,bach_two-site_2010}, where the physical system is mapped onto an impurity with a single bath
site. In the context of SOGET, the
density-functional SOET Hamiltonian is the Hamiltonian of an Anderson
dimer~\cite{senjean_local_2017},
\be
\hat{H}^{\rm imp, 2L}(n_0)&\equiv&-t
\sum_{\sigma}(\hat{a}^\dagger_{0\sigma}\hat{a}_{1\sigma} +
\hat{a}^\dagger_{1\sigma}\hat{a}_{0\sigma})+U\hat{n}_{0\uparrow}\hat{n}_{0\downarrow}
\nonumber\\
&&+\Delta v^{\rm emb}(n_0)\Big(\hat{n}_1 -
\hat{n}_0\Big)/2,
\ee
where, according to Eq.~(\ref{eq:vemb_decomp_KS-imp}) and Ref.~\cite{senjean_local_2017}, the embedding
potential can be written as follows:
\be
\Delta v^{\rm emb}(n_{0}) = \Delta v^{\rm KS}(n_{0}) - \Delta v_{\rm Hxc}^{\rm imp}(n_{0})
,
\ee  
with
\be
\Delta v^{\rm KS}(n_{0})=\frac{2t(n_{0}-1)}{\sqrt{n_{0}(2-n_{0})}}
=\dfrac{\partial T_{\rm s}^{2L}(n_0)}{\partial n_0}.
\label{eq:kohn_sham_potential_hubbard_dimer}
\ee
$T_{\rm s}^{2L}(n_0)$ denotes the non-interacting density-functional kinetic energy
and 
\be
\Delta v_{\rm Hxc}^{\rm imp}(n_{0})&=&
-\frac{U}{2}n_{0} - \frac{\partial
E_{\rm c}^{\rm imp,2L}(U,n_{0})}{\partial n_{0}}
\nonumber\\
&=&-\frac{U}{2}n_{0} - \frac{\partial
E_{\rm c}^{\rm 2L}(U/2,     n_{0})}{\partial n_{0}}
,
\label{eq:impurity_potential_anderson_dimer}
\ee
where $E_{\rm c}^{\rm 2L}(U,n_{0})$ is the density-functional
correlation energy of the two-electron Hubbard
dimer~\cite{carrascal_hubbard_2015}, which has been introduced in Eq.~(\ref{eq:two_level_impurity_corr_energy}).
While the occupation $n_0$ of the impurity can fluctuate, the total number of
electrons in the dimer is fixed ($n_1=2-n_0$). The bath is reduced to a single site
which plays the role of a reservoir. As a result, we can shift the
embedding potential by $-\Delta v^{\rm emb}(n_0)/2$, thus leading to the
final expression,
\be\label{eq:final_exp_SOET_Hamil_dimer}
\hat{H}^{\rm imp, 2L}(n_0)&\equiv&-t
\sum_{\sigma}(\hat{a}^\dagger_{0\sigma}\hat{a}_{1\sigma} +
\hat{a}^\dagger_{1\sigma}\hat{a}_{0\sigma})+U\hat{n}_{0\uparrow}\hat{n}_{0\downarrow}
\nonumber\\
&&-\Delta v^{\rm emb}(n_0)
\hat{n}_0.
\ee
The embedding potential on the impurity site can be rewritten as follows:
\be
-\Delta v^{\rm emb}(n_0)&=&-\frac{U}{2}n_{0} -\frac{\partial
E_{\rm c}^{\rm imp,2L}(U,n_{0})}{\partial n_{0}}
\nonumber\\
&&-\dfrac{\partial T_{\rm s}^{2L}(n_0)}{\partial n_0},
\ee
and compared with its expression in the true 
impurity-interacting system [see
Eq.~(\ref{eq:2L-BALDA-mu_minus_vemb})]. We note that, since the two
expressions only differ by non-interacting kinetic energy contributions, it is relevant to
use the 2L model
described in
Eq.~(\ref{eq:final_exp_SOET_Hamil_dimer}) as reference for
extracting a density-functional self-energy, especially when electron
correlation is strong.\\ 
%We chose a zero-valued potential on site in order to align the Green's functions and resulting self-energy around $\omega=0$. 

From the exact expressions in Eqs.~(\ref{eq:vemb_decomp_KS-imp}) and
(\ref{eq:exact_imp_Sigma_exp}), we can construct an (approximate) impurity-interacting
density-functional self-energy within the 2L model,
\be
\Sigma^{\rm
imp,2L}_{\rm Hxc}\left(n_{0},\omega\right)=
-\Delta v_{\rm Hxc}^{\rm imp}(n_{0})
+\Sigma^{\rm
imp,2L}_{\rm c}\left(n_{0},\omega\right),
\label{eq:two_level_impurity_self_energy_expression}
\ee
where the frequency-dependent impurity correlation self-energy is
obtained as follows:
\begin{eqnarray}
\Sigma^{\rm
imp,2L}_{\rm c}\left(n_{0},\omega\right)=
\dfrac{1}{G_{0\sigma,0\sigma}^{\rm KS,2L}(n_{0},\omega)} - \dfrac{1}{G_{0\sigma,0\sigma}^{\rm
imp,2L}(n_{0},\omega)}.
\nonumber\\
\label{eq:local_impurity_correlation_self_energy}
\end{eqnarray}
The analytical derivation of both KS and impurity-interacting Green's functions is detailed in
Appendix~\ref{app:dimer_gf}. Note that, in the symmetric and strongly
correlated limits, the impurity self-energy reduces to the
exact atomic self-energy, the well-known Hubbard-I (H-I) approximation~\cite{hubbard_j._electron_1963},
%\begin{eqnarray}
%& &\Sigma_{{\rm Hxc}}^{\rm imp, 2L}(n_{0}=1,\omega)
%\xrightarrow{U/t\rightarrow\infty} \frac{U}{2} + \frac{(U/2)^{2}}{\omega
%+ \im\eta - \varepsilon_{0} - U/2},
%\nonumber\\
%\end{eqnarray}
\begin{eqnarray}
\Sigma_{{\rm Hxc}}^{\rm imp, 2L}(n_{0}=1,\omega)
&\xrightarrow{U/t\rightarrow\infty}& 
\Sigma_{{\rm Hxc}}^{\rm
H-I}\left(n_{0}=1,\mu=\frac{U}{2},\omega\right),
\nonumber\\
\label{eq:hi_self_energy_half_filling}
\end{eqnarray}
where
\be\label{eq:HI_approx_any_dens}
&&
\Sigma_{{\rm Hxc}}^{\rm
H-I}(n_{0},\mu,\omega)
=\frac{U}{2}n_0
\nonumber\\
&&+\dfrac{n_0}{2}\left(1-\dfrac{n_0}{2}\right)\dfrac{U^2}{\omega+{\rm
i}\eta+\mu-(1-\frac{n_0}{2})U}.
\ee
%where $\varepsilon_{0}=-U/2$.

From now on, the local impurity-interacting self-energy introduced in
Eq.~(\ref{eq:local_approx_SOGET}) will be approximated by the 2L one:
\be
\Sigma^{\rm
imp}_{\rm Hxc}\left(n_{0},\omega\right)\rightarrow \Sigma^{\rm
imp,2L}_{\rm Hxc}\left(n_{0},\omega\right).
\label{eq:two_level_impurity_self_energy}
\ee

\subsection{Choice of the chemical potential}\label{subsec:mu_alignment}

The most straightforward way to implement SOGET consists in solving,
for a {\it fixed} $\mu$ value of the chemical potential, the self-consistent
Eqs.~(\ref{eq:local_impurity_self_energy_0})-(\ref{eq:hybridization_function}) within the
2L-BALDA approximation. The
combination of these
equations leads to the following compact one:
\be\label{eq:fixedmu_soget_compact}
G^{-1}(\omega) &=& \omega +{\rm i}\eta+\mu
-\dfrac{U}{2}n^G_0-
\left.\dfrac{\partial e_{\rm c}^{\rm BA}(n)}{\partial
n}\right|_{n=n^G_0}
\nonumber\\
&&-\Delta\left(\mu,n^G_0,\omega\right)-\Sigma^{\rm
imp,2L}_{\rm c}\left(n^G_{0},\omega\right).
\ee 
Unfortunately, this procedure becomes numerically unstable for large $U/t$
values in the range of $\mu$ values that correspond to the Mott--Hubbard
transition (not shown). This is probably due to the discontinuity of the BALDA
correlation potential at half-filling~\cite{xianlong_lattice_2012}. In the rest of this section, we 
present a simplified implementation of SOGET where the chemical potential $\mu$
is replaced by its BALDA {\it filling}-functional
expression [see Sec.~\ref{subsubsec:chem_pot_balda}]. As a result, an
approximate Green's function $G(\omega)$ (and the corresponding 
impurity occupation $n^G_{0}$) will be determined (semi-)
self-consistently for a given filling 
$N/L$. Note that $n^G_{0}$ may actually
deviate from $N/L$ due to the various density-functional approximations
we use. Let us also stress that our 
simplified (semi-) self-consistent SOGET equation will involve the 
continuous KS
density-functional chemical potential only, not the physical
discontinuous one, both on the impurity site {\it and} in the hybridization
function, thus preventing any convergence 
issues. The same procedure will be employed in our second strategy [see
Sec.~\ref{subsubsec:min_grand_can_ener}] where the filling-functional grand canonical 
SOGET energy is minimized with respect to the filling, for a fixed
chemical potential value $\mu$. 

\subsubsection{Density-functional chemical
potential}\label{subsubsec:chem_pot_balda}

The simplest way to prevent convergence issues in SOGET consists in using the BALDA density-functional
expression for the chemical potential, 
\begin{eqnarray}
\mu\rightarrow \mu^{\rm BA}(n_0) = \mu^{\rm KS}(n_0) +\dfrac{U}{2}n_0+ \dfrac{\partial e^{\rm BA}_{\rm
c}(n_{0})}{\partial n_{0}},
\label{eq:BALDA_chem_pot}
\end{eqnarray}
thus leading to the following
substitution in Eqs.~(\ref{eq:local_impurity_self_energy_0}) and
(\ref{eq:local_weiss_field_green_function}) [see
Eqs.~(\ref{eq:2L-BALDA-mu_minus_vemb}),
(\ref{eq:impurity_potential_anderson_dimer}), and
(\ref{eq:two_level_impurity_self_energy_expression})]:  
\be
\mu -
v_{0}^{\rm emb}(n_{0})-{\Sigma}_{\rm Hxc}^{\rm
imp,2L}({n}_0,\omega)
&\rightarrow& \mu^{\rm KS}(n_0)
\nonumber
\\
&&-\Sigma^{\rm
imp,2L}_{\rm c}\left(n_{0},\omega\right). 
\nonumber\\
\ee
As a result, the self-consistent SOGET equation can be further simplified as
follows:  
\begin{eqnarray}
G^{-1}(\omega) &=& 
\omega +{\rm i}\eta+\mu^{\rm KS}\left(n^G_0\right) 
- \Sigma^{\rm
imp,2L}_{\rm c}\left(n^G_{0},\omega\right)
\nonumber\\
&&
-
\Delta\left(\mu=\mu^{\rm BA}\left(n^G_0\right),n^G_0,\omega\right)
. 
\label{eq:dyson_eq_3}
\end{eqnarray}
A fully-self-consistent optimization [with an updated impurity site
occupation in the hybridization function, as depicted in
Eq.~(\ref{eq:dyson_eq_3})] gives, when it converges, too low occupations,
thus preventing any investigation of the Mott--Hubbard transition, for
example. This problem could only be solved through a semi-self-consistent
optimization of the
impurity site occupation. In this case, the latter is frozen to a given filling
$N/L$ in all density-functional contributions but the
impurity-interacting
correlation self-energy, thus leading to our final simplified SOGET
equation,   
\begin{eqnarray}
&&G^{-1}(\omega) = 
\omega +{\rm i}\eta
- \Sigma^{\rm
imp,2L}_{\rm c}\left(n^G_{0},\omega\right)
\nonumber\\
&&
+\mu^{\rm KS}\left(\dfrac{N}{L}\right) 
-
\Delta\left(
%\mu=\mu^{\rm BA}\left(\dfrac{N}{L}\right),
\dfrac{N}{L},\omega\right)
, 
\label{eq:semi-sc_soget_eq}
\end{eqnarray}
where 
%$\Delta\left(
%\mu=\mu^{\rm BA}\left(\dfrac{N}{L}\right),
the hybridization function is determined from Eqs.~(\ref{eq:hybridization_function}) and
(\ref{eq:simplification_inside_hyb_func}) by setting $n_0=N/L$. 

%\manu{For a given filling N/L, we could use the exact BA chemical
%potential instead of the BALDA one.}\\

\subsubsection{Minimization of the per-site grand canonical
energy}\label{subsubsec:min_grand_can_ener}

Another strategy for investigating the variation of the impurity site
occupation with the chemical potential consists in minimizing grand
canonical SOGET per-site energies. For a given filling $N/L$, we can generate from Eq.~(\ref{eq:semi-sc_soget_eq}) 
a self-consistently converged
local Green's function $G
\left({N}/{L},\omega\right)$. The latter is then used to compute the
the impurity site occupation and the per-site energy, thus providing the grand
canonical per-site energy to be minimized with respect to $N/L$ for a
given $\mu$ value. The procedure can be summarized as follows: 
\be\label{eq:minimizing_filling_given_mu}
\dfrac{N(\mu)}{L}=\argmin_{N/L}\left\lbrace e\Big(G({N}/{L})\Big) -
\mu\,n_0^{G
\left({N}/{L}\right)} \right\rbrace,
\ee
%\begin{eqnarray}
%\min_{N}\lbrace e(n^{G^{\rm imp},N}) - \mu\frac{N}{L} \rbrace 
%\label{eq:min_e_mu_n_2}
%\end{eqnarray}
where $G
\left({N}/{L},\omega\right)$ fulfills 
Eq.~(\ref{eq:semi-sc_soget_eq}) and
\begin{eqnarray}
 e(G) &=& \left[t_{\rm s}(n) + t\frac{\partial e_{c}^{\rm
BA}(n)}{\partial t}\right]_{n=n^G_0} 
+ U d\left(G\right)
\label{eq:per_site_energy_soget}
\end{eqnarray}
is the (2L-BALDA) SOGET per-site energy with physical double occupation 
\be
d(G)=
 \left[\frac{\partial e_{c}^{\rm BA}(n)}{\partial U} 
- \frac{\partial E_{c}^{\rm 2L}(U/2,n
)}{\partial U}\right]_{n=n^G_0}+d^{\rm imp}\left(G\right)
\nonumber\\
\label{eq:d_soget_corrected}
\ee   
and
\begin{eqnarray}
d^{\rm imp}\left(G\right) = -\dfrac{1}{\pi
U}\int_{-\infty}^{0}\ddroit\omega\, {\rm Im}\left[\Sigma_{\rm
Hxc}^{\rm imp,2L}\left(n_0^G,\omega\right)G(\omega)\right]
.
\nonumber\\
\label{eq:dimp_soget}
\end{eqnarray}
The final impurity site occupation value is then determined from the
minimizing filling in Eq.~(\ref{eq:minimizing_filling_given_mu}) as follows: 
\be
n_0(\mu)=n_0^{G\left(N(\mu)/L\right)}.
\label{eq:n_0_wrt_mu_Laurent_scheme}
\ee

\section{Summary and computational details}\label{sec:comp_details}

%%%%%%%%%%%%%%%%%%
%% Fig. 1
%%%%%%%%%%%%%%%%%%
\begin{figure}
\resizebox{0.49\textwidth}{!}{
\includegraphics[scale=1.5]{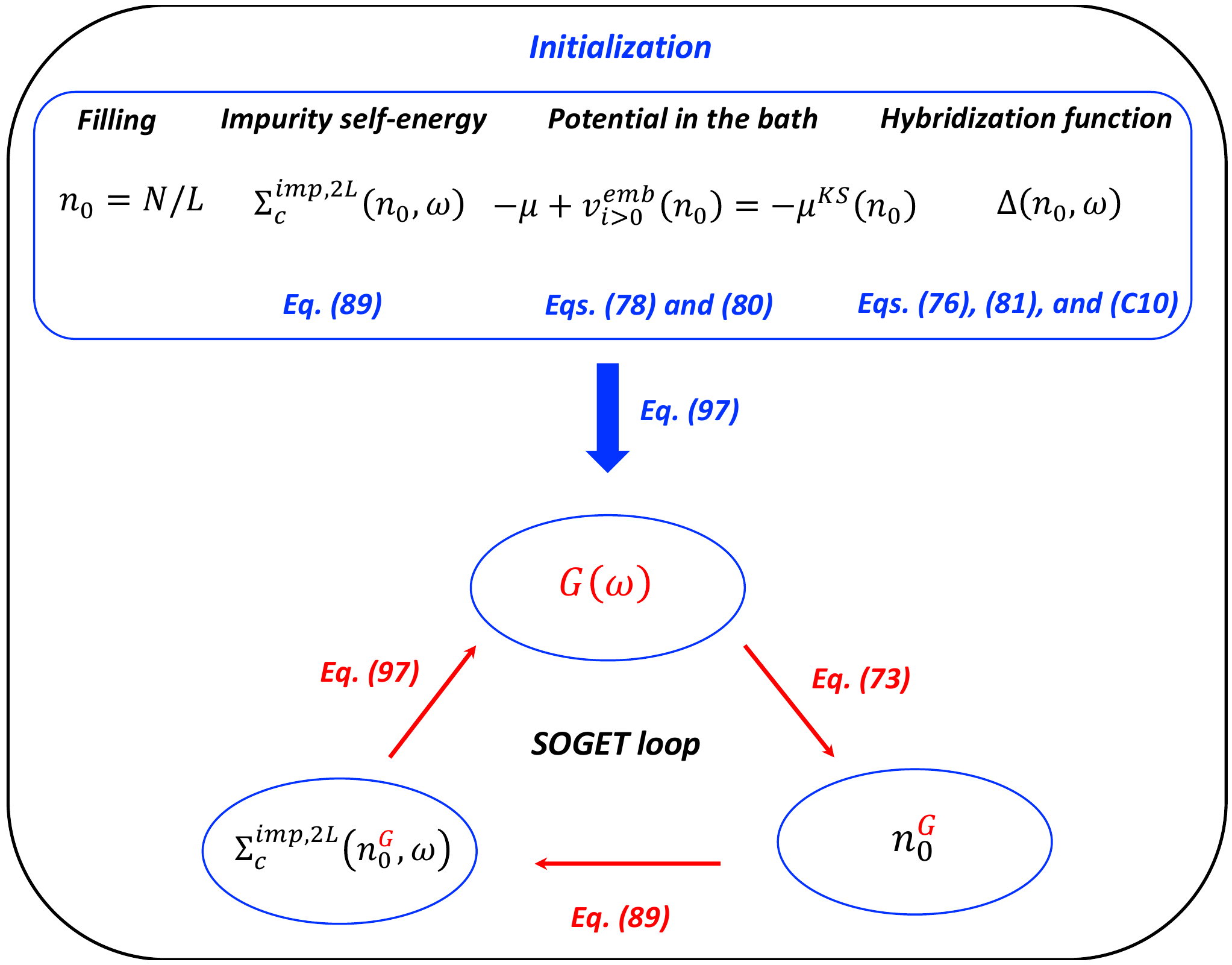}
}
\caption{Schematics of the (semi-) self-consistent implementation of
SOGET used in this work.
%\matt{SOGET Self-Consitent pentagram and relation with other self-consitent approaches, namely, DFT, SOET and DMFT.}
}
%\label{fig:SOGET_diabolic_pentagram}
\label{fig:SOGET_implementation}
\end{figure}
%%%%%%%%%%%%%%%%%%%%%%%
%%% checking Refs. in Fig. 1
%%%%%%%%%%%%%%%%%%%%%%%%%
\iffalse%%%%%
\manu{current ref/updated ref.\\
Eq. 89/\ref{eq:local_impurity_correlation_self_energy}\\
Eq. 78/\ref{eq:chem_pot_0}\\
Eq. 80/\ref{eq:2L-BALDA-mu_minus_vemb_bath}\\
Eq. 76/\ref{eq:hybridization_function}\\
Eq. 81/\ref{eq:simplification_inside_hyb_func}\\
Eq. C10/\ref{eq:V0k_exp}\\
Eq. 97/\ref{eq:semi-sc_soget_eq}\\
Eq. 73/\ref{eq:density_from_converged_G}\\
Eq. 89/\ref{eq:local_impurity_correlation_self_energy}
} 
\fi%%%%%
%%%%%%%%%%%%%%
In order to implement SOGET, we had to make a series of
approximations which have been discussed in detail in
Sec.~\ref{sec:approximations}. A graphical summary of our implementation  
is given in
Fig.~\ref{fig:SOGET_implementation}. The key steps, in both the
initialization and the (semi-) self-consistency cycle of SOGET, are highlighted. 
Density-functional correlation energies have been modelled at the
2L-BALDA level of approximation 
(see
Sec.~\ref{subsec:approx_soet}). 
The method has been applied to the 1D Hubbard model with
$L=400$ sites and a smearing parameter of $\eta=0.01$. As we use an
impurity-interacting self-energy with explicit dependence on the density,
calculations are extremely cheap and not
limited by the size of the system so that, in practice, any filling can
be reproduced. 
Comparison is made with conventional (KS) BALDA and exact BA
results~\cite{lieb_absence_1968,shiba_magnetic_1972}. For analysis
purposes, exact and approximate SOGET spectral functions have been
computed for a half-filled 12-site Hubbard ring {\it via} an exact
diagonalization~\cite{lin_exact_1993} with 100 Lanczos iterations and a
peak broadening of $\eta=0.05$. In all calculations, we set $t=1$.

\section{Results and discussion}\label{sec:res_disc}

\subsection{Self-consistently converged site
occupations}\label{subsec:discuss_sc_occ}

%%%%% Fig. 2 %%%%%%%%%%%%%%%%

\begin{figure}
\resizebox{0.49\textwidth}{!}{
\includegraphics[scale=1]{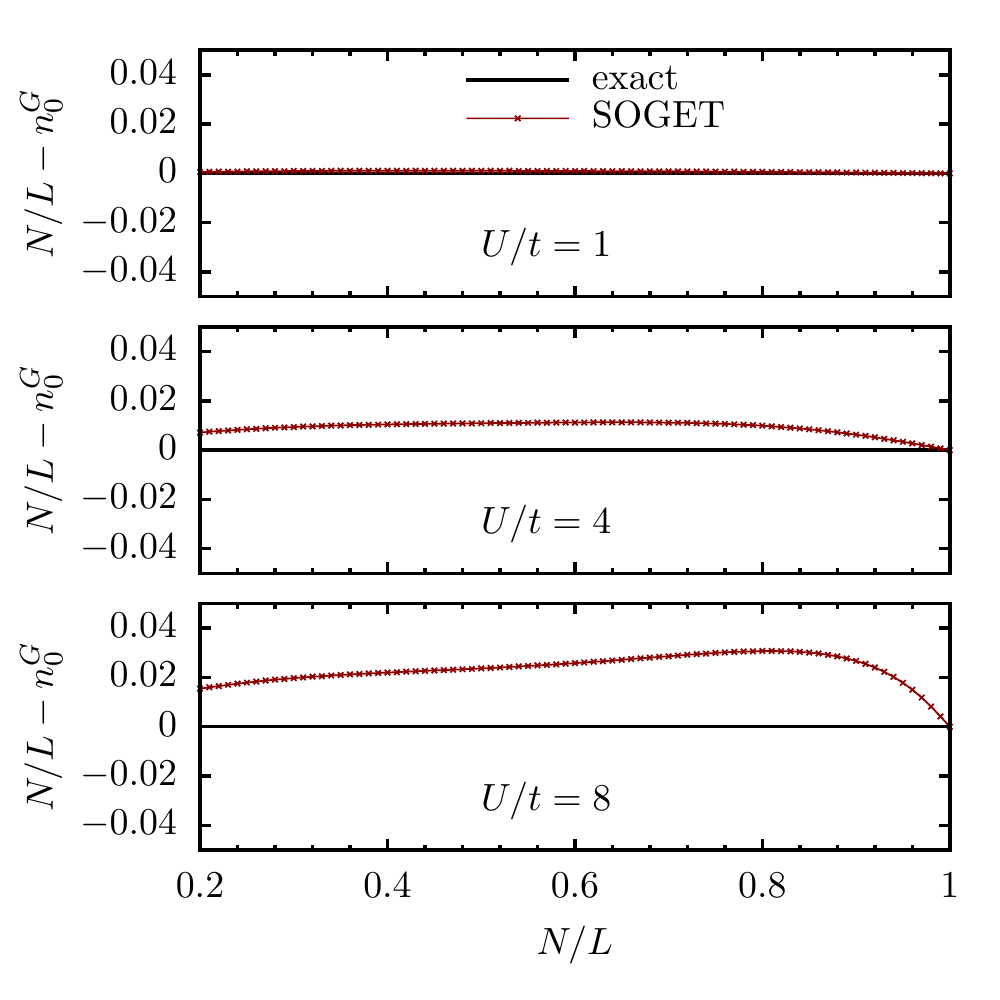}
}
\caption{Deviation of the exact filling $N/L$ [$L=400$] from the
converged impurity site occupation $n^G_{0}$ for different interaction
strengths. The self-consistently converged Green's function and the
corresponding site occupation are computed
according to
Eqs.~(\ref{eq:semi-sc_soget_eq}) and
(\ref{eq:density_from_converged_G}), respectively.}
\label{fig:occupation_error}
\end{figure}

%\red{
%------------------\\
%Fig.~\ref{fig:occupation_error}\\
%------------------\\
%}

Ideally, the self-consistently converged impurity Green's function should restore the exact filling
of the physical Hubbard model. Despite the use of approximate
density-functional (self-) energies, it turns out to be the case at
half-filling (see Fig.~\ref{fig:occupation_error}), as expected from Sec.~\ref{subsec:local-Sigma}. 
In the hole-doped case, however, the converged impurity site occupation 
deviates from the exact filling. In the weakly correlated regime the
error is almost unnoticeable but it becomes more important as we approach
the strongly correlated regime. The deviation remains relatively small
though, unlike in the standard implementation of
SOET~\cite{senjean_site-occupation_2018,senjean_multiple_2018}. In the latter
case, the computation of a
many-body wavefunction allows for unphysical charge transfers between
impurity and bath sites. Such excitation processes are favored by the
approximations made in the density-functional embedding potential. More
precisely, as shown in
Ref.~\cite{senjean_multiple_2018}, the
deviation of the converged impurity occupation from the exact filling is
controlled by the relative position of the 2L impurity-interacting and BALDA correlation
potentials on the impurity site, which actually changes with 
site occupation. In SOGET, this does not occur as the problem is
fully mapped onto the impurity site by using an hybridization function
{\it and} the BALDA density-functional chemical potential. 
As shown in Fig.~\ref{fig:occupation_error}, the impurity occupation is
systematically lower than the exact filling in all correlation regimes, and the error smoothly vanishes when
approaching half-filling, unlike in SOET (see Fig.~7 of
Ref.~\cite{senjean_multiple_2018}). 

%%%%%%%%%%%%%%%%%%%%%%%%%%%%%%%%%%%%%%%%%%%%%%%%%%%%%%%%
\subsection{Double occupations and per-site energies}

%%%%% Fig. 3 %%%%%%%%%%%%%%%%

\begin{figure}
\resizebox{0.49\textwidth}{!}{
\includegraphics[scale=1]{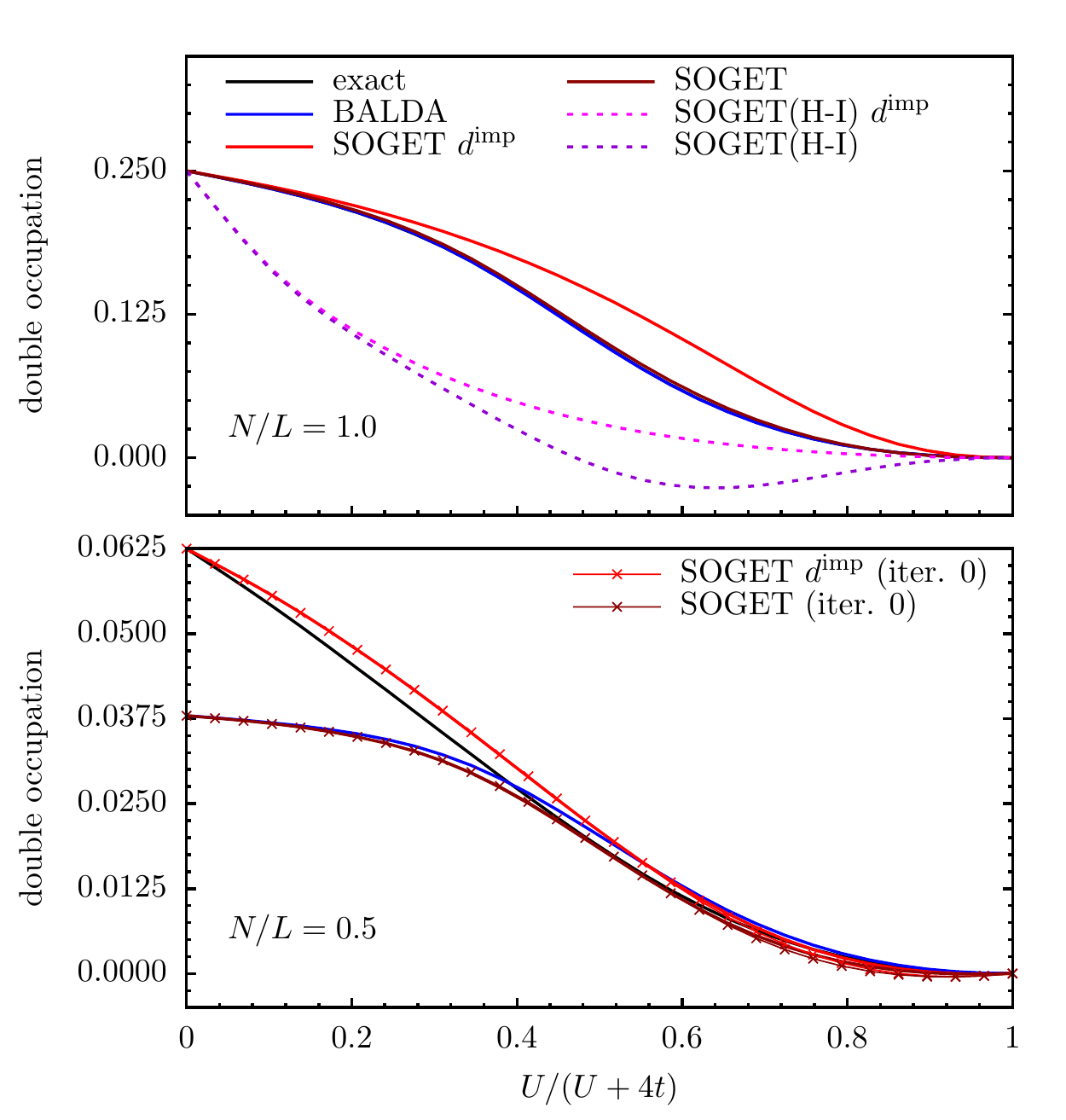}
}
\caption{SOGET double occupation computed as a function of the correlation
strength at half- (top panel) and
quarter-fillings (bottom panel) according to
Eqs.~(\ref{eq:semi-sc_soget_eq}), (\ref{eq:d_soget_corrected}) and
(\ref{eq:dimp_soget}). Comparison is made with conventional BALDA and
exact BA
results (which are equal at half-filling, by construction). Double
occupations obtained without density-functional corrections ($d^{\rm
imp}$) and/or within the Hubbard-I (H-I) self-energy approximation (at half-filling only)
are shown for analysis purposes. Self-consistency effects (which vanish at
half-filling) are slightly   
visible at quarter filling only in the strongly correlated limit [see
``iter. 0'' curves].      
}
\label{fig:double_occ_u}
\end{figure}

%\red{
%------------------\\
%Fig.~\ref{fig:double_occ_u}\\
%------------------\\
%}

Double occupations obtained with and without density-functional
corrections are shown in Fig.~\ref{fig:double_occ_u}. They are plotted
as functions of $U/(U+4t)$ in order to cover all correlation regimes,
from the weakly [$U/(U+4t)\rightarrow 0$] to the strongly correlated
one [$U/(U+4t)\rightarrow 1$]. At half-filling, the bare SOGET impurity
double-occupation obtained from Eq.~(\ref{eq:dimp_soget}) is, like in
SOET~\cite{senjean_site-occupation_2018,senjean_multiple_2018} or DMET~\cite{knizia_density_2012}, too high. While, in DMET, this issue is solved by
increasing the number of impurities, we recover here almost the exact
result with a single impurity by adding the appropriate 
density-functional correction [terms in square brackets on the
right-hand side of Eq.~(\ref{eq:d_soget_corrected})].
Interestingly, the improvement is also substantial in SOET~\cite{senjean_multiple_2018} but not as
impressive as in SOGET. This is due to error cancellations. 
Indeed, combining our approximate local impurity-interacting Green's function with
the 2L impurity-interacting
self-energy leads to an underestimation of
the bare impurity double occupation [see the accurate DMRG values
labelled as ``iBALDA($M$=1)'' in
Fig.~2 of Ref.~\cite{senjean_multiple_2018}]. Consequently, SOGET yields better results than
SOET when the 2L-BALDA density-functional correction is applied.\\

For comparison, we
also computed SOGET double occupations obtained by substituting the H-I
self-energy for the 2L impurity-interacting one. As shown in the top panel of
Fig.~\ref{fig:double_occ_u}, in this case, the bare impurity double occupancy is far
from the physical one except in both non-interacting and
$U/t\rightarrow+\infty$ 
limits. Due to the absence of the hopping parameter $t$ in the atomic
limit, the H-I self-energy overestimates the effect of the Coulomb
interaction $U$ and tends to localize the electrons as soon as $U/t$
deviates from zero. The inclusion of a non-local hopping parameter in
the 2L approximation apparently mimics the fluctuations between the bath
and the impurity and favors the delocalization of electrons. Note that
adding density-functional corrections to the bare impurity double
occupancy deteriorates the results
further when the H-I self-energy is employed. Unphysical negative double
occupations are even obtained in intermediate correlation regimes. H-I performs also poorly away from half-filling (not
shown).\\

\begin{figure}
\resizebox{0.49\textwidth}{!}{
\includegraphics[scale=1]{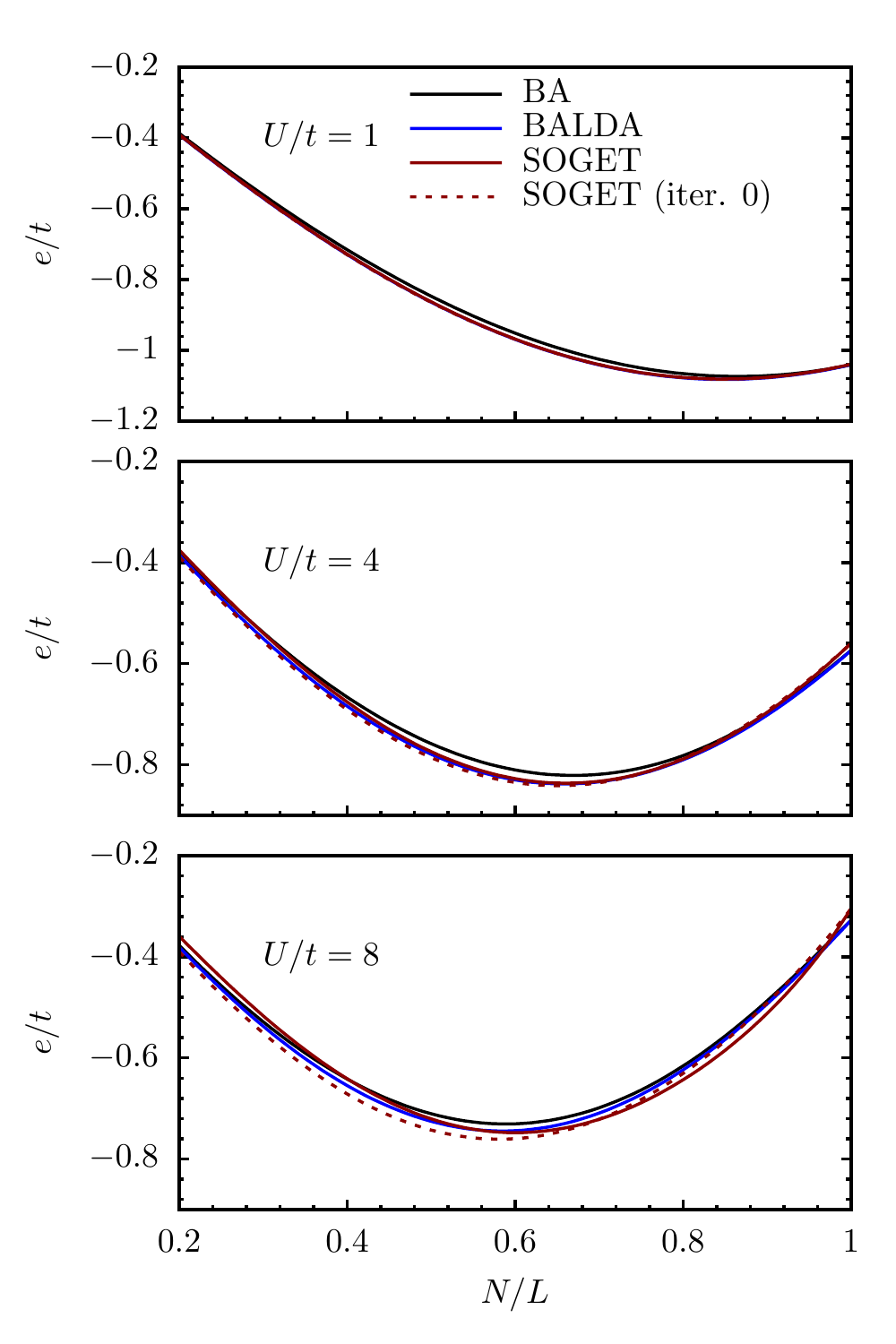}
}
\caption{SOGET per-site energy plotted as a function of the filling
$N/L$ [$L=400$] according to
Eqs.~(\ref{eq:semi-sc_soget_eq}) and (\ref{eq:per_site_energy_soget})
for different correlation strengths. Comparison is made with
conventional BALDA and exact BA results. Results obtained at iteration 0
in Eq.~(\ref{eq:semi-sc_soget_eq})
(i.e. when $n^G_0=N/L$) are also shown [dashed lines] for analysis
purposes. SOGET and BALDA curves are almost indistinguishable for
$U/t=1$.}
\label{fig:persite_energy}
\end{figure}

At quarter-filling, SOGET slightly underestimates the exact double
occupation in the strongly correlated regime [i.e. when $U/(U+4t)>0.5$],
as shown in the bottom panel of Fig.~\ref{fig:double_occ_u}. 
However, in the weakly correlated regime, 
SOGET yields wrong double occupations once the density-functional
corrections are applied. The error is inherited from BALDA [first
term on the right-hand side of Eq.~(\ref{eq:d_soget_corrected})] which, by construction, reproduces the exact BA result only at half-filling.
Away from half-filling, the BALDA correlation functional exhibits an
unphysical linear variation in $U$ (see Eq.~(31)
in Ref.~\cite{senjean_site-occupation_2018}) which artificially lowers the
double occupation in the $U/t\rightarrow 0$ limit.
%%%%%%%%
%n=0.5
%gnuplot> deriv=0.25*(sin(pi*n*0.5)-(n**2))-((n*pi)/8.0)*cos(pi*n*0.5)
%gnuplot> print deriv
%-0.0245633965208121
%gnuplot> print 0.0625-0.0245633965208121
%0.0379366034791879
%%%%%%%%%%%
In this case, the
bare impurity double occupation is much more accurate.
Per-site energies are shown in Fig.~\ref{fig:persite_energy}. For all
fillings and correlation strengths, SOGET yields accurate results
and even improves on previous results from
SOET~\cite{senjean_site-occupation_2018,senjean_multiple_2018}. 
\\
%%%%%%%%%%%%%%%%%%%%%%%%%%%%%%%%%%%%%%%%%%%%%%%%%%%%%%%%

%\begin{figure}
%\resizebox{0.49\textwidth}{!}{
%\includegraphics[scale=1]{../figures/double_occupation_hi}
%}
%\caption{The bare impurity double occupation of SOGET calculated with the 2L and the Hubbard-I self-energy
% [Eq.~(\ref{eq:hi_self_energy_half_filling})] as a function of $U$ at half-filling. The results are compared to the BA. \manu{$L=????$}}
%\label{fig:double_occ_u_hi}
%\end{figure}

%%%%% Fig. 4 %%%%%%%%%%%%%%%%

%\red{
%------------------\\
%Fig.~\ref{fig:persite_energy}\\
%------------------\\
%}

%While
%self-consistency has no significant impact on the double 
%occupations (see the bottom panel of Fig.~\ref{fig:double_occ_u}), it gives 
%slightly more accurate per-site
%energies (especially at lower density) as $U/t$ increases.\\ 
Finally, as for the
comparison of conventional BALDA with SOGET, both approaches qualitatively
exhibit the same performance. 
In the light of
Eqs.~(\ref{eq:d_soget_corrected}) and (\ref{eq:dimp_soget}), we can
conclude 
that the locality of the self-energy, which was assumed in
Eqs.~(\ref{eq:local_approx_SOGET}) and~(\ref{eq:local_impurity_correlation_self_energy}) and is a
key approximation in DMFT, is also
relevant in SOGET. It also means that the local part of the SOGET Green's
function, which incorporates information about the bath through the
hybridization function, can be combined with the self-energy of a simple
system like the Anderson dimer and deliver meaningful results. 
%Quantitatively, SOGET yields slightly better double occupations than BALDA, whereas the per-site energies are more accurately described by BALDA and overestimated by SOGET. 
%Note that, as expected, the Hubbard-I approximation calculated at half-filling deviates very far from the exact per-site energy.
%%%%%%%%%%%%%%%%%%%%%%%%%%%%%%%%%%%%%%%%%%%%%%%%%%%%%%%%
\subsection{Mott--Hubbard transition}\label{subsec:MH-transition}

%%%%% Fig. 5 %%%%%%%%%%%%%%%%

\begin{figure}
\resizebox{0.49\textwidth}{!}{
\includegraphics[scale=1]{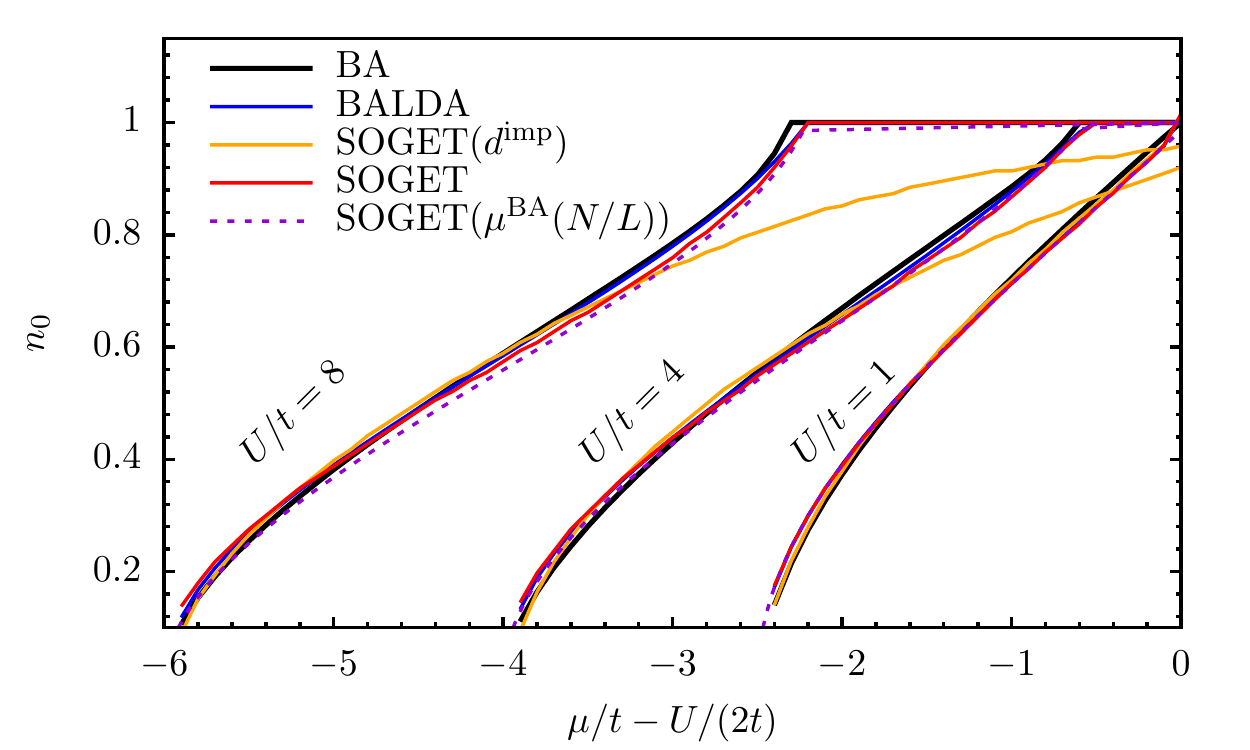}
}
\caption{Mott--Hubbard transition explored in
SOGET for different correlation strengths. In a first approach we plot, as the filling $N/L$
varies, the self-consistently
converged impurity occupation as a function of the BALDA chemical
potential $\mu^{\rm BA}(N/L)$ where $L=400$ [dashed curves]. A second approach
(simply referred to as SOGET) consists in implementing Eqs.(\ref{eq:minimizing_filling_given_mu}) and
(\ref{eq:n_0_wrt_mu_Laurent_scheme}). Results
obtained without density-functional corrections to $d^{\rm imp}$
[SOGET($d^{\rm imp}$)] are also shown. Comparison is made with conventional BALDA and exact BA
results.}
\label{fig:mu_L400}
\end{figure}

%\red{
%------------------\\
%Fig.~\ref{fig:mu_L400}\\
%------------------\\
%}

As discussed in detail in Sec.~\ref{subsec:mu_alignment}, the BALDA chemical potential is
used in SOGET in order to ensure a smooth convergence of the impurity
site occupation in all correlation regimes and fillings. Therefore,
plotting the occupation as a function of the chemical
potential with 
SOGET and conventional BALDA will give exactly the same result if 
self-consistency is neglected. In this case, the Mott--Hubbard
transition is qualitatively well reproduced (see Fig.~\ref{fig:mu_L400}). This well-known feature of
BALDA is due to the derivative discontinuity that the BALDA correlation
potential exhibits at half-filling. In order to evaluate the impact of
the density-functional approximations made in the impurity-interacting correlation
(self-) energy, we first plotted the self-consistently converged
occupation with respect to the filling-functional BALDA chemical
potential. Results are shown in Fig.~\ref{fig:mu_L400}. As expected from
Fig.~\ref{fig:occupation_error}, in
the strongly correlated regime, we observe a slight deviation from
BALDA.\\
\iffalse%%%%%%%%%%% In fact no points between N/L=398/400 and N/L=1 ...
Most importantly, in the region of the Mott--Hubbard
transition, a small but visible increase of the occupation with the
chemical potential appears, thus showing that the opening of the gap has
been lost. This is probably due to the absence of derivative discontinuity
in the 2L impurity potential~\cite{senjean_multiple_2018}.\\
\fi%%%%%%%%%%%%%

Another (less straightforward though) way to
investigate the transition consists in minimizing the SOGET per-site
grand canonical energy according to Eqs.~(\ref{eq:minimizing_filling_given_mu}) and
(\ref{eq:n_0_wrt_mu_Laurent_scheme}). As clearly seen from
Fig.~\ref{fig:mu_L400}, we obtain similar results to BALDA. A
slight deviation appears as the correlation strength increases but the
plateau is relatively well reproduced. Most importantly, if we remove the
density-functional corrections to the ``bare'' impurity double occupation, the
Mott--Hubbard transition disappears. The results look then quite similar to those obtained in
single-site DMET. It clearly shows that BALDA (or, more precisely, the
derivative discontinuity that its correlation potential exhibits at
half-filling) plays a crucial role in the description of the transition
in the single-impurity formulation of SOGET,
as expected from Sec.~\ref{subsubsec:spec_fun_imp_int_system}.\\ 

Note that, once the density-functional
corrections to the double impurity site occupation [see Eq.~(\ref{eq:d_soget_corrected})] are included, we
essentially obtain the right answer through error cancellations. If we
turn to the exact Green's function's expression in Eq.~(\ref{eq:exact_sc_eq_imp-bath}) and
the discussion that follows, the shift in chemical potential observed at
half-filling should, on the impurity site, originate from 
the impurity-interacting self-energy,
while the derivative discontinuities in the fully- and
impurity-interacting correlation potentials should cancel each other.     
These features should of course be reflected in the per-site energy [see the exact
expression given in Eqs.~(\ref{eq:ecbath_exps}) and (\ref{eq:dimp_exact_soget})-(\ref{eq:exact_ener_er_site_exp_GF})] or, more precisely, in its
density-functional derivative from which the chemical potential can in
principle be extracted.   
%%%%%%%%% from the letter
Unlike in the exact theory, the 2L
approximate impurity-interacting correlation potential does not exhibit a derivative
discontinuity at $n_0=1$ for finite $U/t$ values. This is due to the fact
that the two-electron Anderson dimer it originates from is a {\it
closed} system~\cite{senjean_local_2017,senjean_multiple_2018}. One could make the same comment about the 2L
impurity-interacting
density-functional self-energy. Finally, no discontinuity will appear in
our approximate (semi-)
self-consistently converged Green's
function as it is determined from the (continuous) KS chemical
potential [see Eq.~(\ref{eq:semi-sc_soget_eq})]. As a result, 
in our simplified
implementation of SOGET depicted in Eqs.~(\ref{eq:semi-sc_soget_eq}) and
(\ref{eq:minimizing_filling_given_mu})-(\ref{eq:n_0_wrt_mu_Laurent_scheme}), the BALDA
correlation functional [see the first term on the right-hand side of Eq.~(\ref{eq:d_soget_corrected})]
is responsible for the functional derivative discontinuity which makes
the Mott--Hubbard transition possible. This is also the reason why BALDA and SOGET
exhibit exactly the same gap.  
%%%%%

%%%%%%%%%%%%%%%%%%%%%%%%%%%%%%%%%%%%%%%%%%%%%%%%%%%%%%%%
\subsection{Spectral function}\label{subsec:spectr_func}

%%%%% Fig. 6 %%%%%%%%%%%%%%%%
\begin{figure}
\resizebox{0.49\textwidth}{!}{
\includegraphics[scale=1]{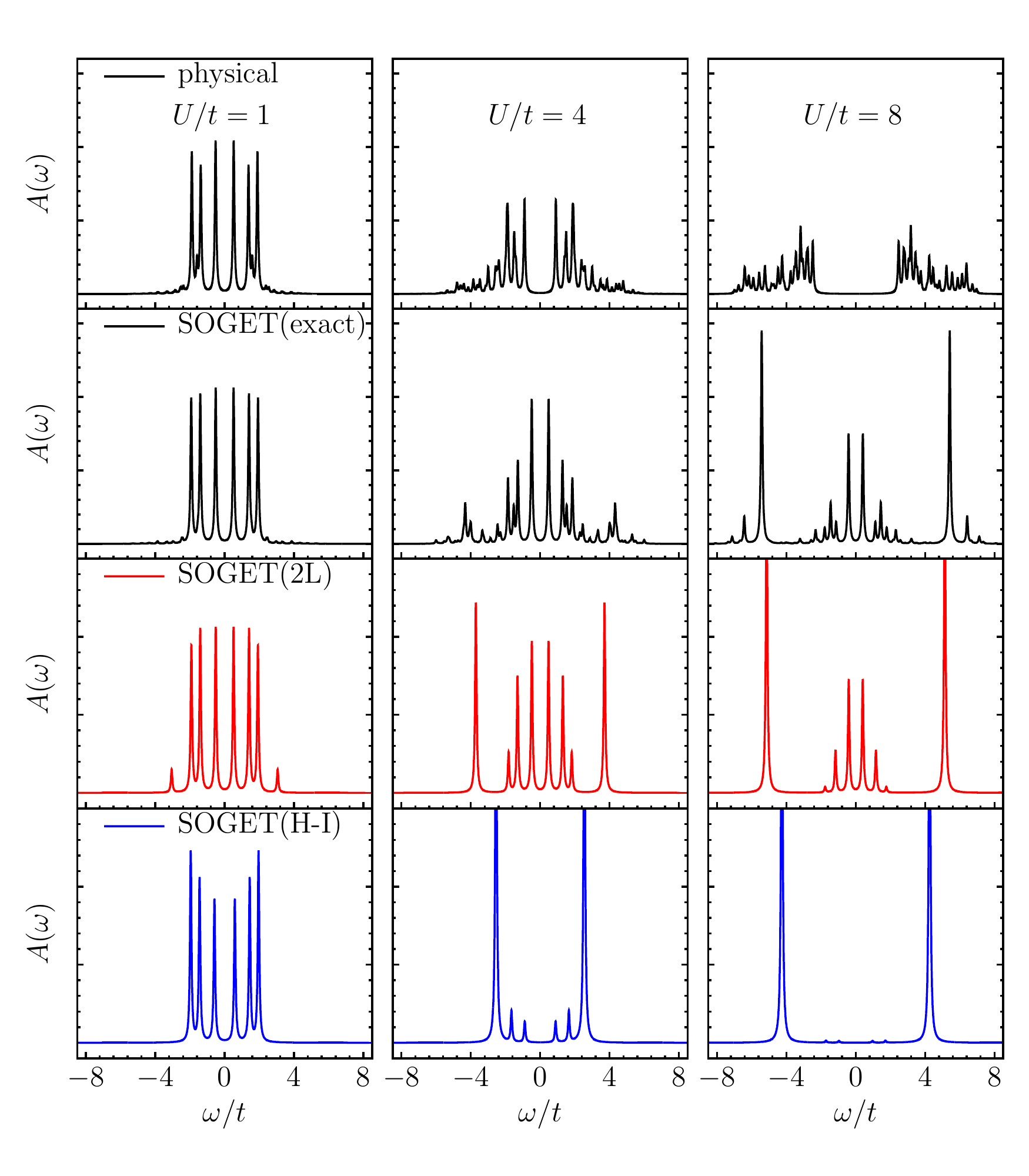}
}
\caption{Spectral functions $A(\omega)=-(1/\pi)\mathrm{Im}\,
\left[G\left(\omega\right)\right]$ computed on the impurity site of a half-filled 12-site
ring ($\mu=U/2$) for various interaction strengths and
Green's functions. From top to bottom:
physical Hubbard model, exact SOGET, SOGET 
with 2L and H-I density-functional self-energies, respectively. See
Sec.~\ref{sec:comp_details} 
for further details.}
%impurity interacting one ${G}_{0\sigma,0\sigma}^{\rm
%imp}\left(\omega\right)$
\label{fig:spectral_functions_L12N12}
\end{figure}

%\red{
%------------------\\
%Fig.~\ref{fig:spectral_functions_L12N12}\\
%------------------\\
%}

%%%%% Fig. 7 %%%%%%%%%%%%%%%%
\begin{figure}
\resizebox{0.49\textwidth}{!}{
\includegraphics[scale=1]{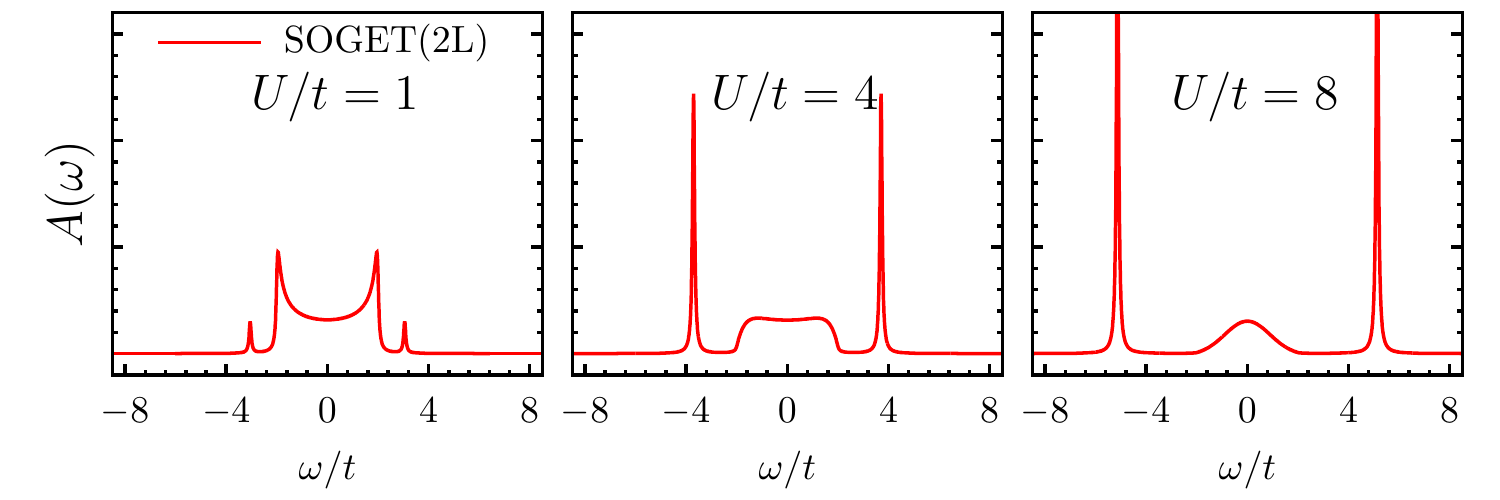}
}
\caption{SOGET spectral function $A(\omega)=-(1/\pi)\mathrm{Im}\,
\left[G\left(\omega\right)\right]$
obtained for a half-filled 400-site
ring and various interaction strengths. The impurity-interacting Green's function
$G\left(\omega\right)$ has been computed according to Eq.~(\ref{eq:semi-sc_soget_eq}) with the 2L
density-functional correlation self-energy. 
}
%impurity interacting one ${G}_{0\sigma,0\sigma}^{\rm
%imp}\left(\omega\right)$
\label{fig:spectral_functions_L400N400}
\end{figure}

Let us now focus on the local Green's function that is computed in
SOGET. For analysis purposes, we first generated the
corresponding
spectral function for a half-filled 12-site ring. Results are shown in
Fig.~\ref{fig:spectral_functions_L12N12} and compared with exact and
other approximate spectral functions. The exact physical and impurity-interacting spectral functions differ
substantially as the correlation strength increases. Indeed, in exact SOET or SOGET, the impurity-interacting
system is expected to reproduce the physical impurity site occupation
only. Like in KS-DFT, one should not expect the physical local Green's function
or its spectral function to be reproduced. The opening of the gap that the physical
Green's function exhibits cannot be seen in the impurity-interacting
system simply because, in this case, the interactions in the bath have
been replaced by a local potential. As discussed in detail in
Sec.~\ref{subsubsec:spec_fun_imp_int_system}, the fact that the latter
potential exhibits a derivative discontinuity in the bath plays a
crucial role in the (effective) description of the physical gap opening. The gap will open when plotting the impurity site occupation
as a function of $\mu$ even though the (impurity-interacting and
bath-non-interacting) spectral gap is closed. 
\\
     
Interestingly, 2L-BALDA reproduces very well many features of the
exact impurity-interacting spectral function, especially in weak and strong
correlation regimes. Some features like satellites are 
missing though, due to the (oversimplified) Anderson-dimer-based 2L
self-energy we use. The spectral function obtained at the same level of
approximation for the half-filled
400-site ring is shown in Fig.~\ref{fig:spectral_functions_L400N400}.
Some features of the 4-site DMET and cluster DMFT spectral functions
shown in Fig.~1 of Ref.~\cite{booth_spectral_2015} are recovered but,
most importantly, the latter exhibit an open gap while our
single-impurity SOGET spectral function does not. Once again, this is
not in contradiction with the gap opening that SOGET exhibits in
Fig.~\ref{fig:mu_L400}, simply because the physical and
impurity-interacting spectral gaps are not expected to match, even in the
exact theory (see Sec.~\ref{subsubsec:spec_fun_imp_int_system}).
It would be interesting to
see how the spectral gap of the impurity-interacting system evolves with the
number of impurities. This would require new developments that are left
for future work.
\\

Finally, if we return to the simpler 12-site model of
Fig.~\ref{fig:spectral_functions_L12N12}, the SOGET spectral function
generated from the Hubbard-I self-energy is much closer to the exact physical
one than the exact impurity-interacting one. In a theory like DMFT where
the local Green's function is the quantity to be reproduced by an
impurity-interacting system, Hubbard-I is a sound approximation.
However, in the light of the double occupation plots shown in
Fig.~\ref{fig:double_occ_u}, the 2L self-energy seems to be a better
choice in SOGET.\\

%%%%%%%%%%%%%%%%%%%%%%%%%%%%%%%%%%%%%%%%%%%%%%%%%%%%%%%%

\section{Conclusions and perspectives}\label{sec:conc_persp}

A novel and in-principle-exact reformulation of SOET (referred to as
SOGET) in terms of
Green's functions has been derived. Once the local self-energy
approximation is made, SOGET becomes formally very similar to DMFT. However,
unlike in DMFT, self-consistency occurs through the density, which is
the basic variable in SOGET. In other words, the impurity-interacting
self-energy is treated as a functional of the (ground-state) density. A
simple 
density-functional approximation based on the Anderson dimer has
been successfully applied to the 1D Hubbard model. While
previous implementations of SOET required the computation of a
correlated many-body
wavefunction for the full impurity-interacting system, SOGET remaps the
impurity correlation problem onto a density-functional dimer. The
drastic reduction in computational cost allowed us to approach the
thermodynamic limit and to model the density-driven Mott--Hubbard
transition (in 1D). Interestingly, thanks to error cancellations, SOGET gave, 
at
half-filling, even more
accurate per-site energies and double occupations than SOET. Spectral functions have also been analyzed. Unlike in
DMFT, the proper description of gap openings in single-impurity SOGET relies on derivative discontinuities in correlation
potentials, like in DFT. The BALDA functional, which was used for
modelling the bath, contains
such a discontinuity,  
by construction. The Mott--Hubbard transition is lost if the latter is
neglected in the SOGET energy expression.\\
%{\it Furthermore,
%Unlike in regular Green's function theory ... in SOGET the density-functional self-energy is uniquely defined by the
%ground state wave function. This is not the case for a self-energy which
%is a functional of the interacting Green's function. In that case the
%functional dependence is multi-valued and a self-consistent calculation
%can lead to nonphysical solutions~\cite{kozik_nonexistence_2015}.
%}

The single-impurity formulation of SOGET presented in this work should
be applicable to the two- and three-dimensional Hubbard model. One key
ingredient, that was missing in the literature until very recently, is
the extension of the one-dimensional BALDA functional to
higher dimensions~\cite{vilela_approximate_2019}. In order to establish
clearer connections between SOGET and DMFT, the infinite-dimension
limit of SOGET should also be explored. Work is currently in progress in
these directions. Note also that, like DFT, SOGET can formally be extended to time-dependent regimes
and finite temperatures, thus giving in principle access to dynamical
properties. The exploration of such extensions is left for future
work.\\

The applicability of SOGET to a wider range of strongly correlated
systems (including {\it ab initio} ones) relies on the development of
density-functional approximations for the (static) impurity-interacting correlation
functional {\it and}
the (dynamical) impurity-interacting correlation
self-energy. The impurity-interacting Sham--Schl\"{u}ter
Eq.~(\ref{eq:imp_SSE}) is a formally-exact constraint which might be
used to develop better approximations to the embedding potential,
provided that we can obtain better density-functional self-energies.
While an explicit density-dependence (like in the 2L Anderson
model) is difficult, if not impossible, to reach for any system,
designing an impurity-interacting self-energy which is an implicit functional of the
density is computationally more demanding but still affordable, in
particular if the size of the system to be described with Green's
functions can be substantially reduced. It would then become possible,
in practice, to extend SOGET to multiple
impurities, like in SOET~\cite{senjean_multiple_2018}. Starting from the SOET
self-consistent Eq.~(\ref{eq:sc_soet_equation}), a simple solution would
consist in applying the Schmidt decomposition to the full
impurity-interacting system and adding dynamical fluctuation corrections,
in the spirit of Refs.~\cite{senjean_projected_2019} and~\cite{fertitta_rigorous_2018}. The latter are expected to be relatively
small due to the density-functional description of the SOET bath. As the
number of impurities increases, we expect the description of
derivative discontinuities in the bath to be less critical, thus making
the accuracy of the method less density-functional-dependent. In order
to turn SOGET into a practical computational method, one should
obviously find the right balance in terms of computational cost (in solving the
impurity-interacting problem) versus feasibility in designing good bath
correlation
functionals.\\
       
Let us mention that an alternative approach, where no many-body wavefunction 
for the full
SOET system would be
needed, would consist in
applying a Householder transformation to the one-electron reduced
density matrix (or, eventually, the frequency-dependent Green's
function) in order to map the properties of the impurity-interacting
system of SOET onto a (much smaller and possibly open) cluster, in the
spirit of
DMET. Work is currently in progress in this direction.\\

Turning finally to 
{\it ab initio} extensions of SOGET, various strategies that recently
appeared in the literature might be considered. The first one is the
Requist--Gross interacting lattice model that is rigorously coupled to
DFT~\cite{requist_model_2019}. Another one is DFT with domain
separation, as proposed by Mosquera
\etal~\cite{JPCA19_Ratner_domain_separation_dft},
which can be seen as an {\it ab initio} generalization of SOET.
%. Their approach relies
%on the separation of real space into domains and is formally exact.
Substituting a Green's function treatment of a given domain for the
many-body wavefunction one would provide an {\it ab initio}
version of SOGET. The latter would in principle be free from double counting,
unlike DMFT+DFT~\cite{kotliar_electronic_2006}.

\section{Acknowledgments}

L.~M. and E.~F. would like to thank B. Senjean for fruitful discussions
and for providing his implementation
of the impurity-interacting correlation potential and energy for the Anderson dimer. They
also thank
M. Tsuchiizu for his useful advices concerning the formalism of Green's functions.
This work was funded by the Ecole Doctorale des Sciences Chimiques 222 (Strasbourg).

%\section{Appendix}

\appendix

\iffalse%%%%%%%%% We just refer to Potthoff's paper %%%%%%%%%%%
{\color{red}
\section{One to one relation between density profile and impurity Green's function}\label{app:one-to-one}
A one to one relation between  $\mathbb{N}_{\rm n}$ and $\mathbb{S}$ that can be establish using:
\begin{equation}
\label{eq:gamma_G}
 {\bf n} = \frac{1}{\beta} \int e^{\imath \beta \omega 0^+}  {\bf G}(\omega),
\end{equation}
so that a surjection exists between the ensemble on {\it v}-representable density and the ensemble of {\it v}-representable self-energy. The second side of the relation is demonstrated  in a more general case where the external potential may be non-local, using the Taylor development at high-frequency of the Green function:
\begin{eqnarray}
\label{eq:G_gamma}
 {\bf G}({\bf r},{\bf r'},\omega) &=& \delta_{{\bf r},{\bf r'}} \frac{1}{\omega} + \frac{1}{\omega^2} V_{\rm ext}({\bf r},{\bf r'}) \nonumber \\ 
 & &+ \frac{1}{\omega^2} \iint \, n({\bf r''},{\bf r'''}) [ F({\bf r},{\bf r''},{\bf r'},{\bf r'''}) \nonumber \\ 
& & - F({\bf r},{\bf r''},{\bf r'''},{\bf r'}) ] {\rm d}{\bf r''}{\rm d}{\bf r'''}  \nonumber \\
 & & + \mathcal{O}(\omega^{-3})
\end{eqnarray}
showing that for a given external potential, two different reduced densities matrices ${\bf n}_1$ and ${\bf n}_2$  lead to two different Green functions. This demonstration holds of course  when the external potential is local, condition for which only the diagonal part of the reduced density matrix has to be considered.}
\fi%%%%%%%%%

\section{Green's function of the Hubbard dimer}\label{app:dimer_gf}

In this section, we calculate the exact Green's function of the singlet ground-state of the asymmetric two-electron Hubbard dimer. We use the following Hamiltonian,
\begin{eqnarray}
\hat{H} = -t
\sum_{\sigma}(\hat{a}^\dagger_{0\sigma}\hat{a}_{1\sigma} +
\hat{a}^\dagger_{1\sigma}\hat{a}_{0\sigma})+\sum^{1}_{i=0}U_{i}\hat{n}_{i\uparrow}\hat{n}_{i\downarrow}
+\sum^{1}_{i=0}v_{i}\hat{n}_{i} \, .
\nonumber\\
\end{eqnarray}
For such as system, the matrix elements of the frequency-dependent retarded Green's function in the Lehmann representation read
\begin{eqnarray}
G_{i\sigma,j\sigma'}(\omega) =
\sum_{a=0}^{1}\dfrac{L_{i\sigma,j\sigma'}^{a}}{\omega + I_{a} + \im\eta}
+ \sum_{b=0}^{1}\dfrac{M_{i\sigma,j\sigma'}^{b}}{\omega + A_{b} + \im\eta}
\nonumber\\
\label{eq:app_dimer_gf}
\end{eqnarray}
where
\begin{eqnarray}
L_{i\sigma,j\sigma'}^{a} = f_{i\sigma,a}f_{j\sigma',a}^{\star}
\end{eqnarray}
and
\begin{eqnarray}
M_{i\sigma,j\sigma'}^{b} = f_{i\sigma,b}f_{j\sigma',b}^{\star}
\end{eqnarray}
are the spectral weights and
\begin{eqnarray}
I_{a} = E_{a}^{N=1} - E_{0}^{N=2}
\end{eqnarray}
and
\begin{eqnarray}
A_{b} = E_{0}^{N=2} - E_{b}^{N=3}
\end{eqnarray}
are the poles of the Green's function. The spectral weights are calculated via the Dyson orbitals defined as follows
\begin{eqnarray}
f_{i\sigma,a} =
\langle\Psi_{0}^{N=2} |\hat{a}_{i\sigma}^{\dagger}|\Psi_{a}^{N=1}\rangle
\end{eqnarray}
and
\begin{eqnarray}
f_{i\sigma,b} =
\langle\Psi_{0}^{N=2} |\hat{a}_{i\sigma}|\Psi_{b}^{N=3}\rangle \, .
\end{eqnarray}
The summations run over the full space of one- and three-electron states of the system. The poles and Dyson orbitals can all be calculated analytically in the case of the Hubbard dimer. First, we solve the trivial one- and three-electron Hubbard dimers. Note that, in the absence of a magnetic field, the Hamiltonians for the two doublets ($s=+1/2$ and $s=-1/2$) are the same. The Hilbert space in the site basis for the one-electron Hubbard dimer reads
\begin{eqnarray}
|\phi_{1}\rangle &=& \hat{a}_{0\uparrow}^{\dagger}|{\rm vac}\rangle
\nonumber\\
|\phi_{2}\rangle &=& \hat{a}_{1\uparrow}^{\dagger}|{\rm vac}\rangle
\nonumber\\
|\phi_{3}\rangle &=& \hat{a}_{0\downarrow}^{\dagger}|{\rm vac}\rangle
\nonumber \\
|\phi_{4}\rangle &=& \hat{a}_{1\downarrow}^{\dagger}|{\rm vac}\rangle \,
\end{eqnarray}
and the Hamiltonian for both doublets ($s=+1/2$ and $s=-1/2$) becomes
\begin{eqnarray}
\mathbf{H}^{N=1} =
\begin{bmatrix}
\begin{array}{cc}
v_{0} & -t  \\
-t & v_{1}  \\
\end{array}
\end{bmatrix} \, .
\end{eqnarray}
The eigenvalues and the corresponding eigenvectors of the one-electron  ($N=1$) Hubbard dimer read
\begin{eqnarray}
E_{0}^{N=1} &=& \dfrac{1}{2}\Big[v_{1} + v_{0} - \sqrt{4t^2 + \Delta v^2}\Big]
\nonumber\\
E_{1}^{N=1} &=& \dfrac{1}{2}\Big[v_{1} + v_{0} + \sqrt{4t^2 + \Delta v^2}\Big]
\nonumber\\
\label{eq:app_energy_one}
\end{eqnarray}
where $\Delta v = v_{1} - v_{0}$, and 
\begin{eqnarray}
c_{0k}^{N=1} = \cos\big(\tan^{-1}(\alpha_{k})\big)
\nonumber\\
c_{1k}^{N=1} = \sin\big(\tan^{-1}(\alpha_{k})\big)
\end{eqnarray}
with $\alpha_{k} = (v_{0} - E^{N=1}_{k})/t$. The wavefunctions of the one-electron Hubbard dimer are expressed as follows,
\begin{eqnarray}
|\Psi_{0}^{N=1} \rangle = c_{00}^{N=1} |\phi_{1}\rangle   + c_{10}^{N=1} |\phi_{2}\rangle 
\nonumber\\
|\Psi_{1}^{N=1} \rangle = c_{01}^{N=1} |\phi_{3}\rangle   + c_{11}^{N=1} |\phi_{4}\rangle 
\end{eqnarray}
\begin{eqnarray}
|\Psi_{2}^{N=1} \rangle = c_{00}^{N=1} |\phi_{1}\rangle   + c_{10}^{N=1} |\phi_{2}\rangle 
\nonumber\\
|\Psi_{3}^{N=1} \rangle = c_{01}^{N=1} |\phi_{3}\rangle   + c_{11}^{N=1} |\phi_{4}\rangle 
\label{eq:app_one_electron_wf}
\end{eqnarray}

The Hilbert space in the site basis for the three-electron ($N=3$) Hubbard dimer reads
\begin{eqnarray}
|\phi_{1}\rangle &=& \hat{a}_{0\uparrow}^{\dagger}\hat{a}_{0\downarrow}^{\dagger}\hat{a}_{1\downarrow}^{\dagger}|{\rm vac}\rangle
\nonumber\\
|\phi_{2}\rangle &=& \hat{a}_{1\uparrow}^{\dagger}\hat{a}_{0\uparrow}^{\dagger}\hat{a}_{1\downarrow}^{\dagger}|{\rm vac}\rangle
\nonumber\\
|\phi_{3}\rangle &=& \hat{a}_{0\uparrow}^{\dagger}\hat{a}_{0\downarrow}^{\dagger}\hat{a}_{1\downarrow}^{\dagger}|{\rm vac}\rangle
\nonumber\\
|\phi_{4}\rangle &=& \hat{a}_{1\uparrow}^{\dagger}\hat{a}_{0\downarrow}^{\dagger}\hat{a}_{1\downarrow}^{\dagger}|{\rm vac}\rangle
\end{eqnarray}
and the Hamiltonian for both doublets ($s=+1/2$ and $s=-1/2$) becomes
\begin{eqnarray}
\mathbf{H}^{N=3} =
\begin{bmatrix}
\begin{array}{cc}
2v_{0} + v_{1} + U_{0}  & -t \\
-t & v_{0} + 2v_{1} + U_{1} \\
\end{array}
\end{bmatrix} \, .
\end{eqnarray}
The eigenvalues and the corresponding eigenvectors of the three-electron Hubbard dimer read
\begin{eqnarray}
E_{0}^{N=3} &=& \frac{1}{2}\Bigg[ U_{0} + U_{1} + 3(v_{0} + v_{1})  
\nonumber\\
& &- \sqrt{4t^{2} + \big(\Delta v + \Delta U\big)^{2} }\Bigg]
\nonumber\\
E_{1}^{N=3} &=& \frac{1}{2}\Bigg[ U_{0} + U_{1} + 3(v_{0} + v_{1}) 
\nonumber\\
& &+ \sqrt{4t^{2} + \big(\Delta v + \Delta U\big)^{2} }\Bigg]
\label{eq:app_energy_three}
\end{eqnarray}
where $\Delta U = U_{1} - U_{0}$, and
\begin{eqnarray}
c_{0k}^{N=3} = \cos\big(\tan^{-1}(\beta_{k})\big)
\nonumber\\
c_{1k}^{N=3} = \sin\big(\tan^{-1}(\beta_{k})\big)
\end{eqnarray}
with $\beta_{k} = (2v_{0} + v_{1} + U_{0} - E_{k})/t $. The wavefunctions of the three-electron Hubbard dimer are expressed as follows,
\begin{eqnarray}
|\Psi_{0}^{N=3} \rangle = c_{00}^{N=3} |\phi_{1}\rangle   + c_{10}^{N=3} |\phi_{2}\rangle 
\nonumber\\
|\Psi_{1}^{N=3} \rangle = c_{01}^{N=3} |\phi_{3}\rangle   + c_{11}^{N=3} |\phi_{4}\rangle 
\end{eqnarray}
\begin{eqnarray}
|\Psi_{2}^{N=3} \rangle = c_{00}^{N=3} |\phi_{1}\rangle   + c_{10}^{N=3} |\phi_{2}\rangle 
\nonumber\\
|\Psi_{3}^{N=3} \rangle = c_{01}^{N=3} |\phi_{3}\rangle   + c_{11}^{N=3} |\phi_{4}\rangle 
\label{eq:app_three_electron_wf}
\end{eqnarray}

Then, we calculate the singlet ground state energy and wavefunction of the two-electron $(N=2)$ Hubbard dimer. We use the following Hilbert space,
\begin{eqnarray}
|\phi_{1}\rangle &=& \hat{a}_{0\uparrow}^{\dagger}\hat{a}_{0\downarrow}^{\dagger}|{\rm vac}\rangle
\nonumber\\
|\phi_{2}\rangle &=& \hat{a}_{1\uparrow}^{\dagger}\hat{a}_{1\downarrow}^{\dagger}|{\rm vac}\rangle
\nonumber\\
|\phi_{3}\rangle &=& \frac{1}{\sqrt{2}}(\hat{a}_{0\uparrow}^{\dagger}\hat{a}_{1\downarrow}^{\dagger} + \hat{a}_{1\uparrow}^{\dagger}\hat{a}_{0\downarrow}^{\dagger})|{\rm vac}\rangle
\end{eqnarray}
The two-electron Hamiltonian then reads
\begin{eqnarray}
\mathbf{H}^{N=2} =
\begin{bmatrix}
\begin{array}{cccc}
U_{0} + 2v_{0} & 0 & -\sqrt{2}t \\
0 & U_{1} + 2v_{1} &  -\sqrt{2}t  \\
-\sqrt{2}t & -\sqrt{2}t & v_{0} + v_{1}   \\
\end{array}
\end{bmatrix} \, .
\end{eqnarray}
The ground-state eigenvalues a solution of the cubic secular equation of the Hamiltonian and is written as follows, 
\begin{eqnarray}
E_{0}^{N=2} = 2\sqrt{-Q}\cos\Big(\frac{\Theta + 2\pi}{3}\Big) + \frac{U_{0}+U_{1}}{3} + v_{0} + v_{1}
\nonumber\\
\label{eq:app_energy_two}
\end{eqnarray}
where
\begin{eqnarray}
\Theta = \cos^{-1}\Big( \frac{R}{\sqrt{-Q^{3}}}\Big) \, ,
\end{eqnarray}
\begin{eqnarray}
R = \dfrac{9a_{2}a_{1} - 27a_{0} - 2a_{2}^{3}}{54}
\end{eqnarray}
and
\begin{eqnarray}
Q = \dfrac{3a_{1} - a_{2}^{2}}{9}
\end{eqnarray}
with
\begin{eqnarray}
a_{0} &=& (v_{0} + v_{1}) (4t^{2} - U_{0} U_{1} - 4 v_{0} v_{1})
\nonumber\\
&+& 2(U_{0} + U_{1}) (t^{2} - v_{0} v_{1}) - 2(U_{0} v_{1}^{2} + U_{1} v_{0}^{2}) \, ,
\nonumber\\
\end{eqnarray}
\begin{eqnarray}
a_{1} &=& U_{0}U_{1} + 8v_{0}v_{1}- 4t^{2} + 2(v_{0}^{2} + v_{1}^{2}) 
\nonumber\\
&+& U_{0}(v_{0} + 3v_{1}) + U_{1}(v_{1} + 3v_{0}) \, ,
\end{eqnarray}
and
\begin{eqnarray}
a_{2} = -(U_{0} + U_{1}) - 3(v_{0} + v_{1}) \, .
\end{eqnarray}

The two-electron wavefunction of the Hubbard dimer reads
\begin{eqnarray}
\Psi_{0}^{N=2} = c_{1}^{N=2}|\phi_{1}\rangle + c_{2}^{N=2}|\phi_{2}\rangle + c_{3}^{N=2}|\phi_{3}\rangle 
\label{eq:app_two_electron_wf}
\end{eqnarray}
where
\begin{eqnarray}
c_{1}^{N=2} = \frac{A}{\sqrt{A^{2}+B^{2} + 2A^{2}B^{2}/(4t^2)}}
\nonumber\\
c_{2}^{N=2} = \frac{B}{\sqrt{A^{2}+B^{2} + 2A^{2}B^{2}/(4t^2)}}
\nonumber\\
c_{3}^{N=2} = \frac{\sqrt{2} \frac{AB}{2t}}{\sqrt{A^{2}+B^{2} + 2A^{2}B^{2}/(4t^2)}}
\end{eqnarray}
with
\begin{eqnarray}
A &=& U_{1} + 2v_{1} - E_{0}^{N=2} 
\nonumber\\
B &=& U_{0} + 2v_{0} - E_{0}^{N=2} \, .
\end{eqnarray}

With Eqs.~(\ref{eq:app_one_electron_wf}), (\ref{eq:app_three_electron_wf}) and {\ref{eq:app_two_electron_wf}), we calculate the Dyson orbitals,
\begin{eqnarray}
f_{0\sigma,a} = 
-\dfrac{A\cos(\tan^{-1}\alpha_{a}) + \frac{AB}{2t}\sin(\tan^{-1}\alpha_{a})}{\sqrt{A^{2} + B^{2} + 2A^{2}B^{2}/(4t^2)}}
\nonumber\\
f_{1\sigma,a} = 
-\dfrac{B\sin(\tan^{-1}\alpha_{a}) +  \frac{AB}{2t}\cos(\tan^{-1}\alpha_{a})}{\sqrt{A^{2} + B^{2} + 2A^{2}B^{2}/(4t^2)}}
\nonumber\\
f_{0\sigma,b} =
\dfrac{B\sin(\tan^{-1}\beta_{b}) +  \frac{AB}{2t}\cos(\tan^{-1}\beta_{b})}{\sqrt{A^{2} + B^{2} + 2A^{2}B^{2}/(4t^2)}}
\nonumber\\
f_{1\sigma,b} =
-\dfrac{A\cos(\tan^{-1}\beta_{b}) +  \frac{AB}{2t}\sin(\tan^{-1}\beta_{b})}{\sqrt{A^{2} + B^{2} + 2A^{2}B^{2}/(4t^2)}}  \, .
\nonumber\\
\end{eqnarray}
The poles are calculated with Eqs.~(\ref{eq:app_energy_one}), (\ref{eq:app_energy_three}) and (\ref{eq:app_energy_two}). Now, we possess all the quantities we need in order to calculate the Green's function [Eq.~\ref{eq:app_dimer_gf}]. The non-interacting Green's function is easily calculated by setting $U_{0}=U_{1}=0$,
\begin{eqnarray}
G^{0}_{0\sigma,0\sigma'}(\omega) &=& \dfrac{\delta_{\sigma\sigma'}}{2}
\bigg[
\dfrac{1+\sin(\tan^{-1}\frac{\Delta v}{2t})}{\omega - (v_{0}+v_{1} - \sqrt{4t^{2} + \Delta v^{2}})/2 + \im\eta}
\nonumber\\
&+& \dfrac{1-\sin(\tan^{-1}\frac{\Delta v}{2t})}{\omega - (v_{0}+v_{1} + \sqrt{4t^{2} + \Delta v^{2}})/2 + \im\eta}
\bigg] \, ,
\nonumber\\
G^{0}_{1\sigma,1\sigma'}(\omega) &=& \dfrac{\delta_{\sigma\sigma'}}{2}
\bigg[
\dfrac{1-\sin(\tan^{-1}\frac{\Delta v}{2t})}{\omega - (v_{0}+v_{1} - \sqrt{4t^{2} + \Delta v^{2}})/2 + \im\eta}
\nonumber\\
&+& \dfrac{1+\sin(\tan^{-1}\frac{\Delta v}{2t})}{\omega - (v_{0}+v_{1} + \sqrt{4t^{2} + \Delta v^{2}})/2 + \im\eta}
\bigg] \, ,
\nonumber\\
G^{0}_{0\sigma,1\sigma'}(\omega) &=& \dfrac{\delta_{\sigma\sigma'}}{2}
\bigg[
\dfrac{\cos(\tan^{-1}\frac{\Delta v}{2t})}{\omega - (v_{0}+v_{1} - \sqrt{4t^{2} + \Delta v^{2}})/2 + \im\eta}
\nonumber\\
&-& \dfrac{\cos(\tan^{-1}\frac{\Delta v}{2t})}{\omega - (v_{0}+v_{1} + \sqrt{4t^{2} + \Delta v^{2}})/2 + \im\eta}
\bigg]
\nonumber\\
\end{eqnarray}
In the non-interacting KS Green's function of the asymmetric dimer $\mathbf{G}^{\rm KS,2L}({n}_0,\omega)$, we choose the following on-site potentials
\begin{eqnarray}
& & v_{0} = -\Delta v^{\rm KS}(n_{0}) = -\frac{2t(n_{0}-1)}{\sqrt{n_{0}(2-n_{0})}}
\nonumber\\
& & v_{1} = 0 \, .
\end{eqnarray}
The interacting impurity Green's function of the asymmetric Anderson dimer  $\mathbf{G}^{\rm imp,2L}({n}_0,\omega)$  as a functional of the site occupation is obtained by switching off the interaction on site 1 ( $U_{0} = U$ and $U_{1} = 0$) and setting
\begin{eqnarray}
& & v_{0} = -\Delta v^{\rm emb}(n_{0}) = -\Delta v^{\rm KS}(n_{0}) + \Delta v_{\rm Hxc}^{\rm imp}(U,n_{0})
\nonumber\\
& & v_{1} = 0 \, ,
\end{eqnarray}
where
\begin{eqnarray}
\Delta v_{\rm Hxc}^{\rm imp}(U,n_{0}) = -\frac{U}{2}n_{0} -
\frac{\partial E_{\rm c}^{\rm 2L}(U/2,n_{0})}{\partial n_{0}}.
\end{eqnarray}
The two-level impurity correlation self-energy [Eq.~(\ref{eq:local_impurity_correlation_self_energy})] is then obtained by inverting the KS and interacting Anderson Green's functions of the impurity site and subtracting the impurity potential $\Delta v_{\rm Hxc}^{\rm imp}(U,n_{0})$,
\begin{eqnarray}
\Sigma_{\rm c}^{\rm imp,2L}({n}_0,\omega) &=& \frac{1}{G_{0\sigma,0\sigma}^{\rm KS,2L}(\omega,n_{0})} - 
 \frac{1}{G_{0\sigma,0\sigma}^{\rm imp,2L}(\omega,n_{0})} 
 \nonumber\\
& &- \Delta v_{\rm Hxc}^{\rm imp}(U,n_{0}) \, .
\end{eqnarray}
The formulas to calculate the exact impurity potential $\Delta v_{\rm Hxc}^{\rm imp}(U,n_{0})$ for the asymmetric Hubbard dimer have been derived in Refs.~\cite{carrascal_hubbard_2015,carrascal_corrigendum:_2016,senjean_multiple_2018}. In the symmetric Anderson dimer ($v_{0} = -U/2$,  $v_{1} = 0$) and $n=1$, there is an analytic expression for the exact impurity Green's function and self-energy~\cite{lange_renormalized_1998},
\begin{eqnarray}
G_{00}^{\rm imp}(\omega) &=& \dfrac{1}{4}
\bigg[
\dfrac{1-\frac{U^{2} - 32t^{2}}{\sqrt{(U^{2}+64t^{2})(U^{2}+16t^{2})}}}{\omega - (\sqrt{U^{2} +64t^{2}} - \sqrt{U^{2} +16t^{2}})/4 + \im\eta}
\nonumber\\
&+& \dfrac{1+\frac{U^{2} - 32t^{2}}{\sqrt{(U^{2}+64t^{2})(U^{2}+16t^{2})}}}{\omega - (\sqrt{U^{2} +64t^{2}} + \sqrt{U^{2} +16t^{2}})/4 + \im\eta}
\nonumber\\
&+& \dfrac{1-\frac{U^{2} - 32t^{2}}{\sqrt{(U^{2}+64t^{2})(U^{2}+16t^{2})}}}{\omega + (\sqrt{U^{2} +64t^{2}} - \sqrt{U^{2} +16t^{2}})/4 + \im\eta}
\nonumber\\
&+& \dfrac{1+\frac{U^{2} - 32t^{2}}{\sqrt{(U^{2}+64t^{2})(U^{2}+16t^{2})}}}{\omega + (\sqrt{U^{2} +64t^{2}} + \sqrt{U^{2} +16t^{2}})/4 + \im\eta}
\bigg] \, ,
\nonumber\\
\end{eqnarray}
\begin{eqnarray}
\Sigma_{{\rm Hxc}}^{\rm imp}(n_{0}=1,\omega) =
\dfrac{U}{2} + \dfrac{1}{2}\bigg[ \dfrac{(\frac{U}{2})^{2}}{\omega - 3t + \im\eta}  + \dfrac{(\frac{U}{2})^{2}}{\omega + 3t + \im\eta} \bigg]  \, .
\nonumber\\
\end{eqnarray}

\section{Hybridization function}\label{app:hybridization_function}

The
approximate interaction-free Green's function on the impurity site  
$\mathcal{G}^{\rm
imp}\left({n}_0,\omega\right)$ can be seen as a
density-functional Weiss field. According to   
Eqs.~(\ref{eq:v-depdt-weiss_field_green_function})
and (\ref{eq:approx_embedding_potential}), it can be expressed as
follows: 
%is an approximation to $\left[\mathbfcal{G}_{\mathbf{v}=0}^{\rm
%imp}\left(\mathbf{n},\omega\right)\right]_{0\sigma,0\sigma}$ 
\be
\mathcal{G}^{\rm
imp}\left({n}_0,\omega\right)=
%\left[
\mathcal{G}_{0\sigma,0\sigma}^{\rm
imp}\left({n}_0,\omega\right)
%\right]
,
\nonumber\\
\ee
where
\be\label{eq:Weiss_field_00comp}
\left[\mathbfcal{G}^{\rm
imp}\left({n}_0,\omega\right)\right]^{-1}=
\left(\omega+\mu+{\rm
i}\eta\right)\mathbf{I}-\mathbf{t}-
\mathbf{v}^{\rm emb}(n_0)
,
\nonumber\\
\ee
and
\be
v^{\rm emb}_0(n_0)&=&\dfrac{\partial e_{\rm c}^{\rm BA}(n_0)}{\partial
n_0}-\dfrac{\partial E^{\rm imp}_{\rm c}(n_{0})}{\partial
n_{0}}, 
\nonumber\\
v^{\rm emb}_i(n_0)&\overset{i>0}{=}&\dfrac{U}{2}n_{0}+\dfrac{\partial e_{\rm c}^{\rm BA}(n_0)}{\partial
n_0}. 
\ee
By diagonalizing the interaction-free Hamiltonian in the bath, 
\be
&&\hat{T}+\sum_{i}v^{\rm emb}_i(n_0)\hat{n}_i=v^{\rm
emb}_0(n_0)\hat{n}_0
\nonumber\\
&&+\sum^{\rm
bath}_k\sum_{\sigma}\varepsilon_k(n_0)\hat{a}^\dagger_{k\sigma}\hat{a}_{k\sigma}
\nonumber\\
&&+\sum^{\rm
bath}_k\sum_{\sigma}\Big(V_{0k}\hat{a}^\dagger_{0\sigma}\hat{a}_{k\sigma}+V^*_{0k}\hat{a}^\dagger_{k\sigma}\hat{a}_{0\sigma}\Big),
\ee
we can rewrite Eq.~(\ref{eq:Weiss_field_00comp}) as follows:
\be
\left[\mathbfcal{G}^{\rm
imp}\left({n}_0,\omega\right)\right]^{-1}=\left[\mathbfcal{G}^{\rm
bath}\left({n}_0,\omega\right)\right]^{-1}-\mathbf{V}
,
\ee
or, equivalently,
\be\label{eq:non-int_Dyson_eq}
\mathbfcal{G}^{\rm
imp}\left({n}_0,\omega\right)&=&\mathbfcal{G}^{\rm
bath}\left({n}_0,\omega\right)
\nonumber\\
&&+\mathbfcal{G}^{\rm
bath}\left({n}_0,\omega\right)\mathbf{V}\mathbfcal{G}^{\rm
imp}\left({n}_0,\omega\right),
\ee
where the matrix elements of $\mathbf{V}$ are
\be
V_{0\sigma,0\sigma'}&=&0,
\nonumber\\
V_{0\sigma,k\sigma'}&=&\delta_{\sigma\sigma'}V_{0k},
\nonumber\\
V_{k\sigma,0\sigma'}&=&\delta_{\sigma\sigma'}V^*_{0k}
\nonumber\\
V_{k\sigma,k'\sigma'}&=&0
\ee
and
\be
\mathcal{G}^{\rm
bath}_{0\sigma,0\sigma'}\left({n}_0,\omega\right)&=&\dfrac{\delta_{\sigma\sigma'}}{\omega+\mu+{\rm
i}\eta-v^{\rm emb}_0(n_0)},
\nonumber\\
\mathcal{G}^{\rm
bath}_{0\sigma,k\sigma'}\left({n}_0,\omega\right)
&=&\mathcal{G}^{\rm
bath}_{k\sigma,0\sigma'}\left({n}_0,\omega\right)=0,
\nonumber\\
\mathcal{G}^{\rm
bath}_{k\sigma,k'\sigma'}\left({n}_0,\omega\right)&=&\dfrac{\delta_{kk'}\delta_{\sigma\sigma'}}{\omega+\mu+{\rm
i}\eta-\varepsilon_k(n_0)}.
\ee

Since, according to Eq.~(\ref{eq:non-int_Dyson_eq}),
\be
&&\mathcal{G}^{\rm
imp}\left({n}_0,\omega\right)=\mathcal{G}^{\rm
bath}_{0\sigma,0\sigma}\left({n}_0,\omega\right)
\nonumber\\
&&+
\mathcal{G}^{\rm
bath}_{0\sigma,0\sigma}\left({n}_0,\omega\right)\sum_kV_{0k}\mathcal{G}_{k\sigma,0\sigma}^{\rm
imp}\left({n}_0,\omega\right),
\ee
and
\be
\mathcal{G}_{k\sigma,0\sigma}^{\rm
imp}\left({n}_0,\omega\right)=
\mathcal{G}^{\rm
bath}_{k\sigma,k\sigma}\left({n}_0,\omega\right)V^*_{0k}\mathcal{G}^{\rm
imp}\left({n}_0,\omega\right),
\ee
we finally obtain
\be
&&\mathcal{G}^{\rm
imp}\left({n}_0,\omega\right)=\mathcal{G}^{\rm
bath}_{0\sigma,0\sigma}\left({n}_0,\omega\right)
\nonumber\\
&&+
\mathcal{G}^{\rm
bath}_{0\sigma,0\sigma}\left({n}_0,\omega\right)\left(\sum_{k}\left\vert
V_{0k}\right\vert^2\mathcal{G}^{\rm
bath}_{k\sigma,k\sigma}\left({n}_0,\omega\right)\right)\mathcal{G}^{\rm
imp}\left({n}_0,\omega\right),
\nonumber\\
\ee 
thus leading to the expression in
Eq.~(\ref{eq:local_weiss_field_green_function}) where
$\Delta(n_{0},\omega)=\sum_{k}\left\vert
V_{0k}\right\vert^2\mathcal{G}^{\rm
bath}_{k\sigma,k\sigma}\left({n}_0,\omega\right)$.

\section{Diagonalization of the interaction-free Hamiltonian in the
bath}\label{appendix:diag_bath}

We divide the non-interacting Hamiltonian into a single impurity and a bath,
\begin{eqnarray}
\mathbf{H}_{0} =
\begin{bmatrix}
\begin{array}{ll}
v_{0}^{\rm emb}(n_{0}) & \mathbf{h}_{\rm 0k} \\
\mathbf{h}_{\rm k0}^{\dagger} & \mathbf{H}_{\rm kk}
\end{array}
\end{bmatrix} - \mu\mathbf{I} \, .
\end{eqnarray}
The block matrix $\mathbf{H}_{\rm kk}$ describes a non-interacting chain connected to a single impurity on both ends. The part $\mathbf{h}_{\rm 0k}$ ($\mathbf{h}_{\rm k0}^{\dagger}$) contains the connection between the impurity and a non-interacting chain through the hopping parameter. We define a unitary transformation that only diagonalizes the bath,
\begin{eqnarray}
\mathbf{U} =
\begin{bmatrix}
\begin{array}{ll}
1 & \mathbf{0} \\
\mathbf{0} & \mathbf{C}
\end{array}
\end{bmatrix}
\end{eqnarray}
where $\mathbf{C}$ contains the eigenvectors of the bath. The transformed Hamiltonian
\begin{eqnarray}
\mathbf{U}^{\dagger}\mathbf{H}_{0}\mathbf{U} =
\begin{bmatrix}
\begin{array}{ll}
v_{0}^{\rm emb}(n_{0}) & \mathbf{V} \\
\mathbf{V}^{\dagger} & \mathbf{E}
\end{array}
\end{bmatrix} 
= \mathbf{H}_{0}^{\rm And}
\end{eqnarray}
is the non-interacting Anderson model, where $\mathbf{V} = \mathbf{h}_{\rm 0k}\mathbf{C}$ are bath-impurity couplings and $\mathbf{E} = \mathbf{C}^{\dagger}\mathbf{H}_{\rm kk}\mathbf{C}$ is a diagonal matrix containing the eigenvalues [ $\varepsilon_{k}(n_{0})$] of the bath orbitals. The non-interacting Green's function is obtained by solving a set of linear equations:
\begin{eqnarray}
\big[ (\omega + \im\eta + \mu)\mathbf{I} - \mathbf{H}_{0}^{\rm And} \big] \boldsymbol{\mathcal{G}}^{\rm imp}(n_{0},\omega) = \mathbf{I}
\end{eqnarray}
In the case of a single impurity, the expressions for the non-interacting Green's function are obtained straightforwardly:
\begin{eqnarray}
\boldsymbol{\mathcal{G}}^{\rm imp}_{00}(\omega,n_{0}) = \dfrac{1}{\omega + \mu + \im\eta  - v_{0}^{\rm emb}(n_{0})  - \Delta (n_{0},\omega)}
\end{eqnarray}
where
\begin{eqnarray}
\Delta (\omega,n_{0}) = \sum_{k=1} \frac{\vert V_{0k}\vert^{2}}{\omega  + \mu + \im\eta - \varepsilon_{k}(n_{0})} \, .
\end{eqnarray}
For a periodic one-dimensional model with $L$ sites, NN hopping $t$ and constant on-site potential $v_{i>0}^{\rm emb}(n_{0})$ in the bath, we obtain the following analytical expressions for $\varepsilon_{k}(n_{0})$ and the matrix elements of $\mathbf{C}$,
\begin{eqnarray}\label{eq:epsilon_k_n0}
\varepsilon_{k}(n_{0}) \overset{i>0}{=} v_{i}^{\rm emb}(n_{0})  - 2t\cos(k) \, ,
\end{eqnarray}
\begin{eqnarray}
C_{ik} = \sqrt{\frac{2}{L}}\sin(ik)
\end{eqnarray}
where
\begin{eqnarray}
k = m\frac{\pi}{L}
\end{eqnarray}
where $m=1,\ldots,L-1$. Consequently, the impurity-bath coupling parameters $\mathbf{V}$ read
\begin{eqnarray}\label{eq:V0k_exp}
V_{0k} = -t\big(C_{1k} \mp C_{L-1,k} \big)
\end{eqnarray}
for periodic ($-$) and anti-periodic ($+$) boundary conditions.

%\bibliography{biblio_soget}

%%% from the .bbl file

\newcommand{\Aa}[0]{Aa}
%

%%%%%%

%%%%%%%%%%%%%%%%%%%%%%%%%%%%%%

%%%%%%%%%%%%%%%%% Stuff removed by Manu %%%%%%%%%%%%%%%%%%%%%

%%%%%%%%%%%%%%%%%%%%%%%%%%%%%%%%%%%%%%%%%%%%%%%%%%%%%%%%%%%%%%%%%
\iffalse%%%%%%%%%%
It is constructed $\mathbfcal{G}^{\rm
imp}\left(\mathbf{n},\omega\right)$ is constructed from 
Eq.~(\ref{eq:impurity_green_function_lehmann}) by performing the following
substitutions, 
\be\label{eq:inter-free-imp_Hamil} 
\hat{H}^{\rm imp}(\mathbf{n})&\rightarrow&\hat{\mathcal{H}}^{\rm
imp}(\mathbf{n})= \hat{T} 
+ \sum_{i} v_{i}^{\rm
emb}(\mathbf{n})\,\hat{n}_{i},
\ee
and
\be
\Psi^{\rm imp}({\bf n})&\rightarrow&\Phi^{\rm imp}({\bf n})
\nonumber\\
\mathcal{E}^{\rm imp}(\mathbf{n})&\rightarrow&\mathcal{E}_{\Phi^{\rm imp}({\bf n})}
,
\ee
where the (single-determinantal) wavefunction $\Phi^{\rm imp}({\bf n})$
is the ground state of $\hat{\mathcal{H}}^{\rm
imp}(\mathbf{n})$ with energy $\mathcal{E}_{\Phi^{\rm imp}({\bf n})}$.
\fi%%%%%%%%%%%%%%%%%%%%%%%%%%%%%%%
%%%%%%%%%%%%%%%%%%%%%%%%%%%%%%%%%%%%%%%%%%%%%%%%%%%%%%%%%%%%%%%%%

%%%%%%%%%%%%%%%%%%%%%%%%%%%%%%%%%%%%%%%%%%%%%%%%
%%%%%%%%%%%%%%%%%%%%%%%%%%%%%%%%%%%%%%%%%%%%%%%%
\iffalse%%%%%%%%%%%
We use this ``density-functional'' impurity Green's function to obtain the corresponding self-energy via the Dyson equation:
where
\begin{eqnarray}
\mathbf{G}^{\rm imp,0}[\mathbf{n}]^{-1} = (\omega+\im\eta)\mathbf{I} - \mathbf{T} - \mathbf{v}^{\rm emb}(\mathbf{n})
\nonumber\\
\label{eq:noninteracting_impurity_green_function}
\end{eqnarray}
is the retarded non-interacting impurity Green's function associated with the impurity Hamiltonian of \Eq{eq:impurity_hamiltonian} without the Coulomb interaction on the impurity and $\eta\rightarrow 0^{+}$.
\fi%%%
%%%%%%%%%%%%%%%%%%%%%%%%%%%%%%%%%%%%%%%%%%%%%%%%
%%%%%%%%%%%%%%%%%%%%%%%%%%%%%%%%%%%%%%%%%%%%%%%%

\iffalse%%%%%%%%%%%%%
\be
\mathbf{\Sigma}_{\rm Hxc}^{\rm imp}\left(\mathbf{n}^{\mathbf{G}^{\rm imp}},\omega\right) = 
\left[\mathbfcal{G}_{\mathbf{v}}^{\rm
imp}\left(\mathbf{n}^{\mathbf{G}^{\rm imp}},\omega\right)\right]^{-1}
- \left[\mathbf{G}^{\rm imp}\left(\omega\right)\right]^{-1}
\nonumber\\
\ee
\fi%%%%%%%%%%%%%%%%%%%%%%%%%%%%%%

\iffalse%%%%%%%%%%%%%%%%%%%%%%%%%%%%%%%%%%%%%
\begin{eqnarray}
\mathcal{G}^{\rm imp,0}[\mathbf{n}]^{-1} = (\omega+\im\eta+\mu)\mathbf{I} - \mathbf{T} - \mathbf{v}^{\rm emb}(\mathbf{n})
\nonumber\\
\end{eqnarray}
\fi%%%%%%%%%%%%%%%%%%%%%%%%%%%%%%%%%

%%%%%%%%%%%%%%%%%%%%%%%%%%%%%%%%%%%%%%%%%%%%%%%%%%%%%%%%%%%%%

\end{document}